%% file: arxiv/ppph_single.tex
\documentclass[onefignum,onetabnum]{siamart171218}

\input{ex_shared}

\ifpdf
\hypersetup{
  pdftitle={Path Polymatrix Games are PPAD-hard},
  pdfauthor={A. Deligkas, J. Fearnley, and R. Savani}
}
\fi

\usepackage{amsfonts}
\usepackage{amsmath}
\usepackage{amssymb}
\usepackage{float}

\usepackage{cmap}
\usepackage{hyperref}

\usepackage{wrapfig}
\usepackage{caption}
\usepackage{subcaption}

\usepackage{booktabs} 
\usepackage{setspace}

\input macros.tex

\usepackage{tikz}
\tikzstyle{state}=[minimum size=1.1cm,inner sep=0cm,draw,circle,node distance=2cm]

\usepackage{xspace}

\newcommand{\myparagraph}[1]{\smallskip \noindent \textbf{#1}}

\newcommand{\eps}{\ensuremath{\epsilon}}

\newcommand{\sbf}{\ensuremath{\mathbf{s}}\xspace}
\newcommand{\pbf}{\ensuremath{\mathbf{p}}\xspace}

\newcommand{\bx}{\ensuremath{\bar{x}}\xspace}

\newcommand{\badd}{\ensuremath{+^b}\xspace}
\newcommand{\bsub}{\ensuremath{-^b}\xspace}
\newcommand{\bmul}{\ensuremath{*^b}\xspace}

\newcommand{\badda}{\ensuremath{\texttt{+}^b}\xspace}
\newcommand{\bsuba}{\ensuremath{\texttt{-}^b}\xspace}
\newcommand{\bmula}{\ensuremath{\texttt{*}^b}\xspace}

\newcommand{\inp}{\ensuremath{\texttt{in}}\xspace}
\newcommand{\outp}{\ensuremath{\texttt{out}}\xspace}
\newcommand{\disp}{\ensuremath{\texttt{disp}}\xspace}
\DeclareMathOperator{\live}{live}
\DeclareMathOperator{\packed}{packed}

\newcommand{\FIXP}{\ensuremath{\mathtt{FIXP}}\xspace}

\newcommand{\linFIXP}{\ensuremath{\mathtt{LinearFIXP}}\xspace}
\newcommand{\twoDlinFIXP}{\texttt{2D-LinearFIXP}\xspace}
\newcommand{\kDlinFIXP}{\texttt{kD-LinearFIXP}\xspace}

\newcommand{\twoDBrouwer}{\texttt{2D-Brouwer}\xspace}
\newcommand{\discreteBrouwer}{\texttt{DiscreteBrouwer}\xspace}
\newcommand{\etDiscreteBrouwer}{\texttt{$\eps$-ThickDisBrouwer}\xspace}

\newcommand{\PPAD}{\ensuremath{\mathtt{PPAD}}\xspace}

\newcommand{\PP}{\ensuremath{\mathtt{P}}\xspace}

\newcount\Comments  
\definecolor{darkgreen}{rgb}{0,0.6,0}
\newcommand{\kibitz}[2]{\ifnum\Comments=1{\color{#1}{#2}}\fi}

\usepackage{todonotes}

\usepackage[algo2e,ruled]{algorithm2e}
 \AtBeginDocument{
 
 }

\usepackage{xcolor}

\begin{document}

\maketitle

\begin{abstract}
We prove that it is \PPAD-hard to compute a Nash equilibrium
in a tree polymatrix game with at most twenty actions per player, and path
polymatrix game with at most four hundred actions per player. 
These are the first \PPAD hardness results for a game with a constant number of 
actions per player where the interaction graph is acyclic.
Along the way we show \PPAD-hardness for finding an $\eps$-fixed
point of a \twoDlinFIXP instance, when $\eps$ is any constant less than
$0.5$. This result is tight, since finding a 0.5-fixed point in a \twoDlinFIXP instance is trivial.
\end{abstract}



\section{Introduction}
\label{sec:intro}

A \emph{polymatrix game} is a succinctly represented many-player game. The
players are represented by vertices in an \emph{interaction graph}, where each
edge of the graph specifies a two-player game that is to be played by the 
adjacent vertices. Each player picks a \emph{pure strategy}, or {\em action}, and 
then plays that action in all of the edge-games that they are involved with. 
They then receive the
\emph{sum} of the payoffs from each of those games. A \emph{Nash equilibrium}
prescribes a mixed strategy to each player, with the property that no player has
an incentive to unilaterally deviate from their assigned strategy.

Constant-action polymatrix games have played a central role in the study of
equilibrium computation. The classical \PPAD-hardness result for finding Nash
equilibria in bimatrix games~\cite{CDT} uses constant-action polymatrix games as
an intermediate step in the reduction~\cite{CDT,DGP}. Rubinstein~\cite{Rub16} later showed
that there exists a constant $\eps>0$ such that computing an $\eps$-approximate
Nash equilibrium in two-action bipartite polymatrix games is
\PPAD-hard, which was the first result of its kind to give hardness for constant $\eps$.

These hardness results create polymatrix games whose interaction
graphs contain cycles. This has lead researchers to study \emph{acyclic} polymatrix
games, with the hope of finding tractable cases. Littman, Kearns, and Singh~\cite{LKS01}
claimed to produce a polynomial-time algorithm for finding a Nash equilibrium in
a two-action tree \emph{graphical game}, where graphical games are a
slight generalization of polymatrix games.
However, their algorithm does not work, which was pointed out by Elkind, Goldberg, 
and Goldberg~\cite{EGG}, who also showed that the natural fix gives an exponential-time
algorithm. 

Elkind, Goldberg, and Goldberg~\cite{EGG} also 
show that a Nash equilibrium can be found in polynomial time for two-action
graphical games whose interaction graphs contain only paths and cycles. They
also show that finding a Nash equilibrium is \PPAD-hard when the interaction
graph has pathwidth at most four.
Later work of~\cite{BLP} provided a QPTAS for
constant-action tree polymatrix games, and then Ortiz and Irfan~\cite{OI} gave
an FPTAS for this case.
All three papers,~\cite{EGG,BLP,OI}, leave as a main open problem the question
of whether it is possible to find a Nash
equilibrium in a tree polymatrix game in polynomial time.

\myparagraph{Our contribution.}
In this work we show that finding a Nash equilibrium in twenty action tree
polymatrix games is \PPAD-hard, and that finding a Nash equilibrium in four
hundred action tree polymatrix games is \PPAD-hard. Combined with the known
\PPAD containment of polymatrix games~\cite{DGP}, this implies that the problem
is \PPAD-complete. These are the first hardness result for polymatrix (or
graphical) games in which the interaction graph is acyclic, and decisively
closes the open question raised by prior work: tree polymatrix games cannot be
solved in polynomial time unless \PPAD is equal to $\mathtt{P}$, and in fact,
this holds even for path polymatrix games.

Our reduction follows a two-step procedure. We first show hardness for games
played on \emph{caterpillar} graphs (see \cref{fig:structure}) which are tree
games that consist of
a single path with small one-vertex branches affixed to every node. We then
give a generic polynomial-time reduction from caterpillar polymatrix games to
path polymatrix games that preserves the equilibrium structure, which then
proves our \PPAD-hardness result for path games.


Our approach is based upon Mehta's proof that \twoDlinFIXP is
\PPAD-hard~\cite{Meh18}. We show that her reduction can be implemented by a
synchronous arithmetic circuit with \emph{constant width}. We then embed the
constant-width circuit into a caterpillar polymatrix game, where each player in
the game is responsible for simulating all gates at a particular level of the
circuit. This differs from previous hardness results~\cite{DGP,Rub16}, where
each player is responsible for simulating exactly one gate from the circuit.

Along the way, we strengthen Mehta's hardness result for
\linFIXP. She showed \PPAD-hardness for finding an exact fixed point of a
\twoDlinFIXP instance, and an $\eps$-fixed point of a \kDlinFIXP instance, where
$\eps$ is polynomially small. 
As part of our reduction to polymatrix games, we study a constrained version of  \twoDlinFIXP in which each gate is required to output numbers that are within $[0, 1]$, and we show that this problem is \PPAD-hard to approximate when $\eps$ is any constant less than $(\sqrt{2} - 1)/2 \approx 0.2071$. When one lifts the restriction of all gates outputting numbers in $[0, 1]$, and therefore obtains the original \twoDlinFIXP problem, 
a straightforward modification of our hardness result gives \PPAD-hardness for all $\eps < 0.5$. This is a tight result for \twoDlinFIXP, because a $0.5$-fixed point can be found trivially.



\myparagraph{\bf Related work.} The class 
\PPAD was defined by Papadimitriou~\cite{Pap94}.
Years later, Daskalakis, Goldberg, and Papadimitriou~\cite{DGP} (DGP) proved \PPAD-hardness for graphical games and 3-player
normal form games. Chen, Deng, and Teng~\cite{CDT} (CDT) extended this result to
2-player games and proved that there is no FPTAS for the problem unless
$\PPAD=\PP$. The observations made by CDT imply that
DGP's result also holds for polymatrix games with constantly-many actions
(but with cycles in the interaction graph) for an exponentially small $\epsilon$.
Rubinstein~\cite{Rub18} showed that there exists a \emph{constant} $\eps>0$ such that computing an $\eps$-NE in binary-action bipartite polymatrix games is \PPAD-hard (again with cycles in the interaction graph).
Very recently, this constant was significantly improved~\cite{deligkas2024pure}, where it was 
shown that it is $\PPAD$-complete to find a $1/48$-NE of a two-action
polymatrix game, and it is $\PPAD$-complete to find an $\eps$-WSNE of a
two-action polymatrix game for all $\eps < 1/3$. Furthermore, even more recently~\cite{graphical-tight} it was shown that for graphical games, a class more general than polymatrix games, it is $\PPAD$-complete to find a $\eps$-NE of a two-action graphical game for every constant $\eps < 1/2$, and that it is $\PPAD$-complete to find an $\eps$-WSNE of a two-action polymatrix game for all constant $\eps < 1$.

Etessami and Yiannakakis~\cite{EY} defined the classes \FIXP and \linFIXP and
they proved that $\linFIXP=\PPAD$.  Mehta~\cite{Meh18} strengthened these
results by proving that two-dimensional \linFIXP equals \PPAD,
building on the result of Chen and Deng~\cite{CD} who proved that 2D-discrete Brouwer is
\PPAD-hard. 

On the positive side, Cai and Daskalakis~\cite{CD11}, proved that NE can be
efficiently found in polymatrix games where every 2-player game is zero-sum.
Ortiz and Irfan~\cite{OI} and Deligkas, Fearnley, and Savani~\cite{DFS}
produced QPTASs for polymatrix games of bounded treewidth (in addition to the
FPTAS of~\cite{OI} for tree polymatrix games mentioned above). For general
polymatrix games, the only positive result to date is a polynomial-time
algorithm to compute a $(\frac{1}{2}+\delta)$-NE~\cite{DFSS}. Finally, an
empirical study on algorithms for exact and approximate NE in polymatrix games
can be found in~\cite{DFIS}.


\section{Preliminaries}
\label{sec: preliminaries}

\myparagraph{Polymatrix games.}
An $n$-player \emph{polymatrix game} is defined by an undirected
\emph{interaction graph} $G=(V, E)$ with~$n$ vertices, where each vertex
represents a player, and the edges of the graph specify which players interact
with each other. 
Each player in the game has $m$ actions, and each edge $(v, u) \in E$ of
the graph is associated with two $m \times m$ matrices $A^{v,u}$ and $A^{u,v}$
which specify a bimatrix game that is to be played between the two players,
where $A^{v, u}$ specifies the payoffs to player $v$ from their interaction with
player $u$.

Each player in the game selects a single action, and then plays that action in
\emph{all} of the bimatrix games with their neighbours in the graph. Their
payoff is the \emph{sum} of the payoffs that they obtain from each of the
individual bimatrix games.

A \emph{mixed strategy} for player $i$ is a probability distribution over 
the $m$ actions of that player, 
a \emph{strategy profile} is a vector $\sbf = (s_1,s_2,\ldots,s_n)$ where $s_i$
is a mixed strategy for player $i$.
The \emph{vector of expected payoffs} for player $i$ under strategy profile $\sbf$
is
$\pbf_i(\sbf) := \sum_{(i, j) \in E}A^{i,j}s_j$.
The \emph{expected payoff} to player $i$ under $\sbf$ is $s_i \cdot
\pbf_i(\sbf)$.
A strategy profile is a \emph{mixed Nash equilibrium} if $s_i \cdot \pbf_i(\sbf)
= \max \pbf_i(\sbf)$ for all $i$, which means that no player can unilaterally
change their strategy in order to obtain a higher expected payoff. 
In this paper we are interested in the problem of computing a Nash equilibrium
of a \emph{tree} polymatrix game, which is a polymatrix game in which the
interaction graph is a tree.

\myparagraph{Arithmetic circuits.}
For the purposes of this paper, each gate in an arithmetic circuit will operate
only on values that lie in the range $[0, 1]$. In our construction, we will use
four specific gates, called \emph{constant introduction} denoted by $c$,
\emph{bounded addition} denoted by \badd, \emph{bounded subtraction} denoted by
\bsub, and \emph{bounded multiplication by a constant} denoted by $\bmul c$. These gates are
formally defined as follows.
\begin{itemize}
\item $c$ is a gate with no inputs that outputs some fixed constant $c \in [0, 1]$.
\item Given inputs $x, y \in [0,1]$ the gate $x \badd y := \min \left(x+y ,1\right)$.
\item Given inputs $x, y \in [0,1]$ the gate $x \bsub y := \max \left(x-y ,0\right)$.
\item Given an input $x \in [0, 1]$, and a constant $c \ge 0$, the gate $x
\bmul c := \min \left(x * c, 1 \right)$.
\end{itemize}
These gates perform their operation, but also clip the output value so that it
lies in the range~$[0, 1]$. Note that the constant $c$ in the $\bmul c$ gate is
specified as part of the gate. Multiplication of two inputs is not allowed.
 
We will build arithmetic circuits that compute functions of the form $[0,
1]^d \rightarrow [0,1]^d$. A circuit $C = (I, G)$ consists of a set $I = \{
\inp_1, \inp_2, \dots, \inp_d \}$ containing $d$ input nodes, and a set $G = \{
g_1, g_2, \dots, g_k \}$ containing $k$ gates. Each gate $g_i$ has a type from
the set $\{c, \badd, \bsub, \bmul c\}$, and if the gate has one or more inputs,
these are taken from the set $I \cup G$. The connectivity structure of the gates
is required to be a directed acyclic graph. 

The \emph{depth} of a gate, denoted by $d(g)$ is the length of the longest path
from that gate to an input. We will build
\emph{synchronous} circuits, meaning that all gates of the form $g_x = g_y \badd g_z$
satisfy $d(g_x) = 1 + d(g_y) = 1 + d(g_z)$, and likewise for gates of the form
$g_x = g_y \bsub g_z$. 
There are no restrictions on $c$-gates, or $\bmul c$-gates.

The \emph{width} of a particular level $i$ of the circuit is defined to be $w(i)
= | \{ g_j \; : \; d(g_j) = i \} |$, which is the number of gates at that level.
The \emph{width} of a circuit is defined to be $w(C) = \max_{i} w(i)$, which is
the maximum width taken over all the levels of the circuit.

\myparagraph{\bf Straight line programs.}
A convenient way of specifying an arithmetic circuit is to write down a straight
line program (SLP)~\cite{EY}. 
\begin{center}
\scalebox{1}{
\begin{minipage}[c]{0.33\textwidth}
\begin{center}
\begin{algorithm2e}[H]
\texttt{x $\leftarrow$ 0.5}\;
\texttt{z $\leftarrow$ x \badda in1}\;
\texttt{x $\leftarrow$ x \bmula 0.5}\;
\texttt{out1 $\leftarrow$ z \badda x}\;
\caption{Example}
\end{algorithm2e}
\end{center}
\end{minipage}%
\begin{minipage}[c]{0.1\textwidth}
\mbox{}
\end{minipage}%
}
\scalebox{1}{
\begin{minipage}[c]{0.5\textwidth}
\begin{center}
\begin{algorithm2e}[H]
\caption{\texttt{if} and \texttt{for} example}
\DontPrintSemicolon
\texttt{x $\leftarrow$ in1 \bmula 1}\;
\For{\texttt{i in  $\{1, 2, \dots, 10\}$}}{
    \If{\texttt{i is even}}{
        \texttt{x $\leftarrow$ x \badda 0.1}\;
    }
}
\texttt{out1 $\leftarrow$ x \bmula 1}\;
\end{algorithm2e}
\end{center}
\end{minipage}
}
\end{center}

\noindent Each line of an SLP consists of a statement of the form \verb+v+
$\leftarrow$ \verb+op+, where \verb+v+ is a \emph{variable}, and \verb+op+
consists of exactly one arithmetic operation from the set
set $\{c, \badd, \bsub, \bmul c\}$. 
The inputs to the gate can be any variable that is defined before the line, or
one of the inputs to the circuit. We permit variables to be used on the left
hand side in more than one line, which effectively means that we allow variables
to be overwritten.

It is easy to turn an SLP into a circuit. Each line is turned into a gate, and
if variable \verb+v+ is used as the input to gate $g$, then we set the
corresponding input of $g$ to be the gate $g'$ that corresponds to the
line that most recently assigned a value to \verb+v+. SLP 1 
above specifies a circuit with four gates, and the output of the circuit will be
$0.75$ \badda $\inp_1$.

For the sake of brevity, we also allow \texttt{if} statements and \texttt{for}
loops in our SLPs.
These two pieces of syntax can be thought of as macros that help us specify a
straight line program concisely. The arguments to an \texttt{if} statement or a
\texttt{for} loop must be constants that do not depend on the value of any gate
in the circuit. When we turn an SLP into a circuit, we unroll every \texttt{for}
loop the specified number of times, and we resolve every \texttt{if} statement
by deleting the block if the condition does not hold. So the example in SLP 2
produces a circuit with seven gates: two gates correspond to the lines \texttt{x
$\leftarrow$ in1 \bmula 1} and \texttt{out1 $\leftarrow$ x \bmula 1}, while
there are five gates corresponding to the line \texttt{x $\leftarrow$ x \badda
0.1}, since there are five copies of the line remaining after we unroll the loop
and resolve the \texttt{if} statements. The output of the resulting circuit will
be $0.5$ \badda $\inp_1$.

\myparagraph{\bf Liveness of variables and circuit width.}
Our ultimate goal will be to build circuits that have small width. To do this,
we can keep track of the number of variables that are \emph{live} at any one
time in our SLPs. A variable \texttt{v} is live at line $i$ of an SLP if both of
the following conditions are met.
\begin{itemize}
\item There exists a line with index $j \le i$ that assigns a value
to \texttt{v}.
\item There exists a line with index $k \ge i$ that uses the value
assigned to \texttt{v} as an argument.
\end{itemize}
The number of variables that are live at line $i$ is denoted by $\live(i)$, and
the number of variables \emph{used by} an SLP is defined to be $\max_i
\live(i)$, which is the maximum number of variables that are live at any point
in the SLP. 

\begin{lemma}
\label{lem:width}
An SLP that uses $w$ variables can be transformed into a polynomial-size
synchronous circuit of width~$w$.
\end{lemma}

\begin{proof}
The idea is to make each level of the circuit correspond to a line of the SLP.
We assume that all \texttt{for} loops have been unrolled, and that all
\texttt{if} statements have been resolved. Suppose that the resulting SLP has
$k$ lines, and furthermore assume that at each line of the SLP, we have an
indexed list $v_1, v_2, \dots, v_l$ of the variables that are live on each line,
where of course we have $l \le w$. 

We will build a circuit with $k \cdot w$
gates, and will index those gates as $g_{i,j}$, where $1 \le i \le k$ is a line,
and $1 \le j \le w$ is a variable. The idea is that the gate $g_{i,j}$ will
compute the value of the $j$th live variable on line $i$. 
The gate $g_{i,j}$ will be constructed as
follows.
\begin{itemize}
\item If there are fewer than $j$ variables live at line $k$ of the SLP, then
$g_{i,j}$ is a dummy $c$-gate.

\item If line $i$ of the SLP is \texttt{$v_j$ $\leftarrow$ op}, then we define
$g_{i,j} = \mathtt{op}$. If \texttt{op} uses a variable $\mathtt{x}$ as an
input, then by definition, this variable must be live on line~$i-1$, and so we
find the index $j'$ for \texttt{x} on line~$i-1$, and we substitute
$g_{i-1,j'}$ for \texttt{x} in \texttt{op}. We do this for both arguments in
the case where \texttt{op} is \badd or \bsub.

\item If line $i$ of the SLP does not assign a value to $v_j$, then by
definition, the variable must be live on line $i-1$. As before, let $j'$ be the
index of this variable on line $i-1$. We define $g_{i,j} = g_{i-1,j'}
\bmul 1$.
\end{itemize}
It is not difficult to see that this circuit exactly simulates the SLP.
Moreover, by construction, we have $d(g_{i, j}) = i$. Hence,
each level of the circuit has width exactly $w$, and
so the overall width of the circuit is $w$.

\end{proof}
\section{Hardness of \twoDBrouwer}

In this section, we consider the following problem. It is a variant 
of two-dimensional Brouwer that uses only our restricted set of bounded gates.

\begin{definition}[\twoDBrouwer]
Given an arithmetic circuit $F : [0, 1]^2 \rightarrow [0, 1]^2$ using gates from
the set \{c, \badd, \bsub, \bmul c\}, find $x \in [0, 1]^2$ such that $F(x) = x$.
\end{definition}

As a starting point for our reduction, we will show that this problem is
\PPAD-hard. Our proof will follow the work of Mehta~\cite{Meh18}, 
who showed that the closely related \twoDlinFIXP problem is \PPAD-hard. 
There are two differences between \twoDBrouwer and \twoDlinFIXP.
\begin{itemize}
\item In \twoDlinFIXP, 
all internal gates of the circuit take and return values from $\mathbb{R}$
rather than $[0, 1]$.
\item \twoDlinFIXP circuits uses gates from the set $\{c, +, -, * c,
\max, \min\}$,
where none of these gates bound their outputs to be in $[0, 1]$. 
\end{itemize}
\noindent In this section,  we present an altered version of Mehta's reduction,
which will show that finding an $\eps$-solution to \twoDBrouwer is \PPAD-hard for a constant $\eps$.


\myparagraph{\bf Discrete Brouwer.}
The starting point for Mehta's reduction is the two-dimensional discrete Brouwer
problem, which is known to be \PPAD-hard~\cite{CD}. This problem is defined over
a discretization of the unit square $[0, 1]^2$ into a grid of points
$G = \{0, 1/2^n, 2/2^n, \dots, (2^n - 1)/2^n\}^2$. 
The input to the problem is a Boolean circuit $C : G
\rightarrow \{1, 2, 3\}$ the assigns one of three colors to each point. The
coloring will respect the following boundary conditions.
\begin{itemize}
\item We have $C(0, i) = 1$ for all $i \ge 0$.
\item We have $C(i, 0) = 2$ for all $i > 0$.
\item We have $C(\frac{2^n-1}{2^n}, i) = C(i, \frac{2^n-1}{2^n}) = 3$ for all $i > 0$.
\end{itemize}
These conditions can be enforced syntactically by modifying the circuit.
The problem is to find a grid square that is \emph{trichromatic}, meaning that
all three colors appear on one of the four points that define the square.

\begin{definition}[\discreteBrouwer]
Given a Boolean circuit $C$ of the form $C : \{0, 1\}^n \times \{0, 1\}^n \rightarrow \{1, 2,
3\}$ that satisfies the boundary conditions, find 
a pair of values $x, y \in \{0, 1\}^n$ such that, for each color $i \in
\{1,2,3\}$, there exists a point $(x', y')$ with $C(x', y') = i$
where 
 $x' \in
\{x, x + 1\}$ and $y' \in \{y, y+1\}$.
\end{definition}

\begin{figure}
\begin{subfigure}[t]{.5\linewidth}
\centering
\raisebox{0.75cm}{
\scalebox{0.25}{ 
\begin{tikzpicture}[scale=1]

\pgfmathsetmacro{\thick}{2.5}

\fill [color=blue!60] (0,\thick) rectangle (\thick,10);
\draw (0,\thick) rectangle (\thick,10);
\node[white,font=\boldmath,scale=1.5] at (0.5*\thick,5+\thick/2) {\Huge $2$};

\fill [color=green!60!black] (\thick,10-\thick) rectangle (10,10);
\fill [color=green!60!black] (10-\thick,\thick) rectangle (10,10);
\draw (10-\thick,\thick) -- (10-\thick,10-\thick) -- (\thick,10-\thick) -- (\thick,10) -- (10,10) -- (10,\thick);
\node[white,font=\boldmath,scale=1.5] at (10-\thick/2,10-\thick/2) {\Huge $3$};

\fill [color=red!80] (0,0) rectangle (10,\thick);
\draw (0,0) rectangle (10,\thick);
\node[white,font=\boldmath,scale=1.5] at (5,\thick/2) {\Huge $1$};

\node[xshift=0.9cm, scale=1.4] at (10,\thick/2) {\Huge $\Big\} \ \epsilon$};

\node[yshift=0.6cm, scale=1.4] at (\thick/2,10) {\LARGE $\overbrace{\hspace{1.6cm}}$};
\node[yshift=1.5cm, scale=1.4] at (\thick/2,10) {\Huge $\epsilon$};

\end{tikzpicture}

}}%
\caption{Our stronger boundary conditions.}\label{fig:boundary}
\end{subfigure}%
\begin{subfigure}[t]{.5\linewidth}
\centering
\scalebox{0.4}{
\usetikzlibrary{arrows.meta}
\tikzset{>={Latex[width=8mm,length=3mm]}}

\begin{tikzpicture}[scale=5]

\coordinate (t1) at (-1, {1-sqrt(2)});
\coordinate (t2) at (1, {1-sqrt(2)});
\coordinate (t3) at (0, 1);
\draw[dashed] (t1) -- (t2) -- (t3) -- (t1);

\node[green!60!black,font=\boldmath] at ($(t1) + (-0.075, -0.15)$) {\LARGE $(-1,1-\sqrt{2})\cdot \epsilon$};
\node[blue!60,font=\boldmath] at ($(t2) + (-0.01, -0.15)$) {\LARGE $(1,1-\sqrt{2})\cdot \epsilon$};
\node[red!90,font=\boldmath] at ($(t3) + (0.4,0)$) {\LARGE $(0,1)\cdot \epsilon$};

\node at (0, {1-sqrt(2)-0.15}) {\LARGE $\underbrace{\hspace{4cm}}_{(\sqrt{2}-1)\cdot \epsilon}$};

\pgfmathsetmacro{\ss}{(sqrt(2)-1)};
\coordinate (s1) at ({-\ss}, {\ss}) {};
\coordinate (s2) at ({-\ss}, {-\ss}) {};
\coordinate (s3) at ({\ss}, {-\ss}) {};
\coordinate (s4) at ({\ss}, {\ss}) {};
\draw[thick, dotted] (s1) -- (s2) -- (s3) -- (s4) -- (s1);

\draw[line width=1.5mm, green!60!black, ->] (0,0) -- (t1);
\draw[line width=1.5mm, blue!60, ->] (0,0) -- (t2);
\draw[line width=1.5mm, red!60, ->] (0,0) -- (t3);

\end{tikzpicture}

}
\caption{The mapping from colors to vectors.}
\label{fig:colors}
\end{subfigure}
\caption{Reducing \etDiscreteBrouwer to \twoDBrouwer.}\label{fig:1}
\end{figure}

Our first deviation from Mehta's reduction is to insist on the following
stronger boundary condition, which is shown in \cref{fig:boundary}. 
\begin{itemize}
\item We have $C(i, j) = 1$ for all $i$, and for all $j \le \eps$.

\item We have $C(i, j) = 2$ for all $j > \eps$, and for all $i \le \eps$.

\item We have $C(i, j) = C(j, i) = 3$ for all $i > \eps$, and all $j \ge 1 -
\eps$.
\end{itemize}
The original boundary conditions placed constraints only on the outermost grid
points, while these conditions place constraints on a border of width $\eps$. 
The reason for the thicker boundary is so that a point in the central region can have a displacement of $\eps$ in any direction without hitting the boundary, which is a property that we will utilize in our reduction.
We call this modified problem
\etDiscreteBrouwer, which is the same as \discreteBrouwer,
except that the function is syntactically required to satisfy the new boundary
conditions.

It is not difficult to reduce \discreteBrouwer to \etDiscreteBrouwer in polynomial time. It suffices to increase the number of points in the grid,
and then to embed the original \discreteBrouwer instance into the $[\eps,
1-\eps]^2$ square in the middle of the instance.

\begin{lemma}
\label{lem:thick}
There is a polynomial-time reduction from \discreteBrouwer to\\ 
\etDiscreteBrouwer.
\end{lemma}

\begin{proof}
Suppose that we are given a \discreteBrouwer instance defined by a circuit $C$ over the grid
$G_n = \{0, 1/2^n, 2/2^n, \dots, (2^n - 1)/2^n\}^2$. 
Let $n'$ be an integer such that $2^n/2^{n'} < (1 - 2 \eps)$. We will build an
\etDiscreteBrouwer instance defined by a circuit $C'$ over the grid 
$G_{n'} = \{0, 1/2^{n'}, 2/2^{n'}, \dots, (2^{n'} - 1)/2^{n'}\}^2$.
We will embed the original instance in the center of the new instance, where the
point $(x_0, y_0) = (0.5 - 2^{n-1}/2^{n'}, 0.5 - 2^{n-1}/2^{n'})$ in $G'$ will correspond to the
point $(0, 0)$ in $G$.
We use the following procedure to determine the color of a point $(x, y) \in
G_{n'}$.
\begin{enumerate}
\item If $0 \le x - x_0 \le 2^n$ and $0 \le y - y_0 \le 2^n$, then $C'(x, y) =
C(x - x_0, y - y_0)$.
\item Otherwise, if $x - x_0 < 0$, then $C(x, y) = 1$.
\item Otherwise, if $y - y_0 \le 0$, then $C(x, y) = 2$.
\item Otherwise, $C(x, y) = 3$.
\end{enumerate}
Observe that 
\begin{equation*}
x_0 = 0.5 - \frac{2^{n-1}}{2^{n'}} > 0.5 - \frac{(1 - 2\eps)}{2} = \eps,
\end{equation*}
where the second inequality used the definition of $n'$. Moreover
\begin{equation*}
x_0 + 2^n = 0.5 + \frac{2^{n-1}}{2^{n'}} < 0.5 + \frac{(1 - 2\eps)}{2} = 1 - \eps,
\end{equation*}
where again the second inequality used the definition of $n'$. The same
inequalities hold for $y_0$. Hence, the first step of our procedure perfectly
embeds the original instance into the new instance, while the other steps ensure
that the \etDiscreteBrouwer boundary conditions hold.

Points in the boundary cannot be solutions, because the boundary constraints
ensure that at least one of the three colors will be missing. Hence, every
solution of $C'$ on $G'$ must also be a solution of $C$ on~$G$. 
\end{proof}

\myparagraph{\bf Embedding the grid in $[0,1]^2$.}
We now reduce \etDiscreteBrouwer to \twoDBrouwer.
One of the keys steps of the reduction is to map points from the continuous
space $[0, 1]^2$ to the discrete grid $G$. Specifically, given a point $x \in
[0, 1]$, we would like to determine the $n$ bits that define the integer
$\lfloor x \cdot 2^n \rfloor$. 

Mehta~\cite{Meh18} showed that this mapping from continuous points to discrete points can be
done by a linear arithmetic circuit. Here we give a slightly different
formulation that uses only gates from the set $\{c, \badd,
\bsub, \bmul c\}$. Let $L$ be a fixed constant that will be defined later.

\begin{center}
\begin{minipage}[c]{0.41\textwidth}
\begin{algorithm2e}[H]
\caption{$\mathtt{ExtractBit(x, b)}$}
\label{slp:extract}
\DontPrintSemicolon
\texttt{b $\leftarrow$ 0.5}\;
\texttt{b $\leftarrow$ x \bsuba b}\;
\texttt{b $\leftarrow$ b \bmula L}\;
\end{algorithm2e}
\end{minipage}%
\begin{minipage}[c]{0.59\textwidth}
\begin{algorithm2e}[H]
\caption{$\mathtt{ExtractBits(x, b_1, b_2, \dots, b_n)}$}
\label{slp:extract-multi}
\DontPrintSemicolon
\For{i in $\{1, 2, \dots, n\}$}{
    \texttt{ExtractBit(x, b$_i$)}\;
    \texttt{y\phantom{$_i$} $\leftarrow$ b$_i$ \bmula 0.5}\;
    \texttt{x\phantom{$_i$} $\leftarrow$ x\phantom{$_i$} \bsuba y}\;
    \texttt{x\phantom{$_i$} $\leftarrow$ x\phantom{$_i$} \bmula 2}\;
}
\end{algorithm2e}
\end{minipage}
\end{center}

SLP~\ref{slp:extract} extracts the first bit of the number $x \in [0, 1]$. The first
three lines of the program compute the value $b = (x \bsub 0.5) \bmul L$. 
There are three possibilities. 
\begin{itemize}
\item If $x \le 0.5$, then $b = 0$.
\item If $x \ge 0.5 + 1/L$, then $b = 1$.
\item If $0.5 < x < 0.5 + 1/L$, then $b$ will be some number strictly between
$0$ and $1$. 
\end{itemize}
The first two cases correctly decode the first bit of $x$, and we call these
cases \emph{good decodes}.
We will call the third case a \emph{bad decode}, since the bit has not been
decoded correctly. 


SLP~\ref{slp:extract-multi} extracts the first $n$ bits of $x$, by
extracting each bit in turn, starting with the first bit. The three lines
after each extraction erase the current first bit of $x$, and then multiply $x$
by two, which means that the next extraction will give us the next bit of $x$.
If any of the bit decodes are bad, then this procedure will break, meaning that
we only extract the first $n$ bits of $x$ in the case where all decodes are
good. We say that $x$ is \emph{well-positioned} if the procedure succeeds,
and \emph{poorly-positioned} otherwise. 

\myparagraph{\bf Multiple samples.}
The problem of poorly-positioned points is common in \PPAD-hardness reductions.
Indeed, observe that we cannot define an SLP that always correctly extracts the
first $n$ bits of $x$, since this would be a discontinuous function, and all
gates in our arithmetic circuits compute continuous functions.
As in previous works, this is resolved by taking multiple samples around a given
point. Specifically, for the point $p \in [0, 1]^2$, we sample $k$ points $p_1$,
$p_2$, \dots, $p_{k}$ where
$p_i = p + \left(\frac{i - 1}{(k + 1) \cdot 2^{n+1}}, \frac{i - 1}{(k + 1) \cdot
2^{n+1}} \right).$
Mehta~\cite{Meh18} proved that there exists a setting for $L$ that ensures that there are at
most two points that have poorly positioned coordinates. We have changed several
details, and so we provide our own statement and proof here.
\begin{lemma}
\label{lem:positioned}
If $L = (k+2) \cdot 2^{n+1}$, then at most two of the points $p_1$ through $p_{k}$
have poorly-positioned coordinates.
\end{lemma}

\begin{proof}
Observe that SLP~\ref{slp:extract} produces a bad decode if and only if $x$ is
in the range $[0.5, 0.5 + 1/L)$. 
Since SLP~\ref{slp:extract-multi} extracts $n$ bits, multiplying $x$ by two each
time, it follows that one of the decodes will fail if 
\begin{equation*}
x \in I(a) = \left[\frac{a} {2^n}, \frac{a}{2^n} + \frac{1}{L} \right),
\end{equation*}
for some integer $a$.

Hence, the point $p_i = (p_{i}^1, p_{i}^2)$ has a poorly-positioned coordinate
if there is some integer~$a$
such that $p_{i}^1 \in I(a)$, or $p_{i}^2 \in I(a)$.
For a fixed dimension $j \in \{1, 2\}$, we have two properties.
\begin{itemize}
\item There cannot be two points $p_i$ and $p_{i'}$ such that $p_{i}^j$ and
$p_{i'}^j$ both lie in the same interval $I(a)$. This is because the width of
the interval is 
\begin{equation*}
\frac{1}{L} = \frac{1}{(k+2) \cdot 2^{n+1}} < \frac{1}{(k+1) \cdot 2^{n+1}},
\end{equation*}
where the final term is the defined difference between 
$p_i^j$ and
$p_{i+1}^j$.

\item There cannot be two distinct indices $a$ and $a'$ such that $p_{i}^j \in
I(a)$ and $p_{i'}^j \in I(a')$. This is because the distance between $p_1^j$ and
$p_{k}^j$ is at most
\begin{equation*}
k \cdot \frac{1}{(k+1) \cdot 2^{n+1}} < \frac{1}{2^{n+1}},
\end{equation*}
whereas the distance between any two consecutive intervals $I(a)$ and
$I(a+1)$ is at least
\begin{equation*}
\frac{a+1}{2^n} - \left(\frac{a}{2^n} + \frac{1}{(k+2) \cdot 2^{n+1}} \right) 
= \frac{1}{2^n} - \frac{1}{(k+2) \cdot 2^{n+1}} > \frac{1}{2^{n+1}}.
\end{equation*}

\end{itemize}
From these two facts, it follows that there is at most one point that has a
poorly-positioned coordinate in dimension $j$, so there can be at most two
points that have poorly positioned coordinates.
\end{proof}

\myparagraph{\bf Evaluating a Boolean circuit.}
Once we have decoded the bits for a well-positioned point, we have a sequence of
0/1 variables. It is easy to simulate a Boolean
circuit on these values. 
\begin{itemize}
\item The operator $\lnot \; x$ can be simulated by $1 \bsub x$.
\item The operator $x \lor y$ can be simulated by $x \badd y$.
\item The operator $x \land y$ can be simulated by applying De Morgan's laws
and using $\lor$ and $\lnot$.
\end{itemize}
Recall that $C$ outputs one of three possible colors. We also assume, without loss of generality, that $C$ gives its output as an \emph{one-hot vector}. This means that
there are three Boolean outputs $x_1, x_2, x_3 \in \{0, 1\}^3$ of the circuit.
The color $1$ is represented by the vector $(1, 0, 0)$, the color $2$ is
represented as $(0, 1, 0)$, and color $3$ is represented as $(0, 0, 1)$. If the
simulation is applied to a point with well-positioned coordinates, then the
circuit will output one of these three vectors, while if it is applied to a
point with poorly positioned coordinates, then the circuit will output some
value $x \in [0, 1]^3$ that has no particular meaning. 

\myparagraph{\bf The output.}
The key idea behind the reduction is that each color will be mapped to a
displacement vector, as shown in \cref{fig:colors}. Here we again deviate
from Mehta's reduction, by giving different vectors that will allow us to prove
our approximation lower bound.
\begin{itemize}
\item Color $1$ will be mapped to the vector $(0, 1) \cdot \eps$.
\item Color $2$ will be mapped to the vector $(1, 1 -\sqrt{2}) \cdot \eps$.
\item Color $3$ will be mapped to the vector $(-1, 1 -\sqrt{2}) \cdot \eps$.
\end{itemize}
These are irrational coordinates, but in our proofs we argue that
a suitably good rational approximation of these vectors will suffice.
We average the displacements over the $k$ different sampled points to get
the final output of the circuit.
Suppose that $x_{ij}$ denotes output $i$ from sampled point $j$. 
Our circuit will compute
$$
\disp_x = \sum_{j = 1}^k \frac{(x_{2j} - x_{3j}) \cdot \eps}{k}, \quad
\disp_y = \sum_{j = 1}^k \frac{\left(x_{1j} + (1 - \sqrt{2} ) (x_{2j} + x_{3j}) \right) \cdot \eps}{k}.
$$
Finally, we specify 
$F : [0, 1]^2 \rightarrow [0, 1]^2$ to compute 
$F(x, y) = (x + \disp_x \cdot \eps, y + \disp_y \cdot \eps)$.

\myparagraph{\bf Completing the proof.}
To find an approximate fixed point of $F$, we must find a point where both
$\disp_x$ and $\disp_y$ are close to zero. 
The dotted square in \cref{fig:colors}
shows the set of displacements that satisfy $\| x - (0, 0) \|_\infty \le
(\sqrt{2} - 1) \cdot \eps$, which correspond to the displacements that would be
$(\sqrt{2} - 1) \cdot \eps$-fixed points. 

The idea is that, if we do not sample points of all three colors, then we cannot
produce a displacement that is strictly better than an $(\sqrt{2} - 1) \cdot \eps$-fixed point. For
example, if we only have points of colors 1 and 2, then the displacement will be
some point on the dashed line between the red and blue vectors in
\cref{fig:colors}. This line touches the box of $(\sqrt{2} - 1) \cdot \eps$-fixed points, but
does not enter it. It can be seen that the same property holds for the other
pairs of colors: we specifically chose the displacement vectors in order to
maximize the size of the inscribed square shown in \cref{fig:colors}.

The argument is complicated by the fact that two of our sampled points may have
poorly positioned coordinates, which may drag the displacement towards $(0, 0)$.
However, this effect can be minimized by taking a large number of samples. We show
the following lemma.

\begin{lemma}
\label{lem:ksamples}
Let $\eps' < (\sqrt{2} - 1) \cdot \eps$ be a constant. There is a sufficiently
large constant $k$ such that, if $\| x - F(x) \|_{\infty} < \eps'$, then $x$ is
contained in a trichromatic square.
\end{lemma}

\begin{proof}
We argue that if $\| x - F(x) \|_{\infty} < \eps'/2$, then there exist three
indices $i_1$, $i_2,$ and $i_3$ such that $p_{i_j}$ has well-positioned coordinates,
and that the lower-left corner of the square containing $p_{i_j}$ has color $j$.

Suppose for the sake of contradiction that this is not true. Then there must be
a color that is missing, and there are two cases to consider. 
\begin{enumerate}
\item First suppose that color 1 is missing. Since there are at most two points
with poorly-positioned coordinates,  we know that we have at least
$k - 2$ points $j$ for which $x_{2j} = 1$ or $x_{3j} = 1$. Hence we have
\begin{equation*}
\disp_y \le \left( \frac{(1 - \sqrt{2} ) (k-2)}{k} + \frac{2}{k} \right) \cdot \eps,
\end{equation*}
where the $2/k$ term comes from the fact that the poorly positioned
points can maximize $\disp_y$ by fixing $x_{1j} = 1$ and $x_{2j} = x_{3j} = 0$,
and thus can 
contribute at most $2 \cdot \eps/k$ to the sum.

As $k$ tends to infinity, the right-hand side converges to
$(1 - \sqrt{2}) \cdot \eps$. Since $\eps' < \eps$, we can choose a sufficiently
large constant $k$ such that $\disp_y < (1 - \sqrt{2}) \cdot \eps'$. Now,
observing that $1 - \sqrt{2}$ is negative, we get the following
\begin{equation*}
\| x - F(x) \|_\infty > \left| (1 - \sqrt{2}) \cdot \eps' \right| = (\sqrt{2} - 1) \cdot
\eps',
\end{equation*}
giving our contradiction.

\item Now suppose that one of colors 2 or 3 is missing. We will consider the
case where color 3 is missing, as the other case is symmetric.
As before, since there are at most two points
with poorly-positioned coordinates,  we know that we have at least
$k - 2$ points $j$ for which $x_{1j} = 1$ or $x_{2j} = 1$.
One of the two following cases applies.

\begin{enumerate}
\item At least $(\sqrt{2} - 1) \cdot k - 2$ well-positioned points satisfy
$x_{2j} = 1$. If this is the case, then we have
\begin{equation*}
\disp_x \ge \left( \frac{(\sqrt{2} - 1) \cdot k - 2}{k} - \frac{2}{k} \right) \cdot
\eps, \\
\end{equation*}
where we have used the fact that there are no well positioned points with color
3, and the fact that the poorly-positioned points cannot reduce the sum by more
than $\frac{2 \cdot \eps}{k}$. 

As $k$ tends to infinity, the right-hand side tends to 
$(\sqrt{2} - 1) \cdot \eps$, so there is a sufficiently large constant $k$ such
that $\disp_x > (\sqrt{2} - 1) \cdot \eps'$, and so
$\| x - F(X) \|_\infty > (\sqrt{2} - 1) \cdot \eps'$.

\item At least $k - (\sqrt{2} - 1) \cdot k$ well-positioned points
satisfy $x_{1j} = 1$.
In this case we have
\begin{align*}
\disp_y &\ge 
\sum_{j = 1}^k \left( \frac{x_{1j} - (\sqrt{2} - 1) x_{2j}}{k} - \frac{2}{k}
\right) \cdot \eps \\
&\ge \left( \frac{ \Bigl( k - (\sqrt{2} - 1) \cdot k \Bigr) - \Bigl((\sqrt{2} - 1)(\sqrt{2} -
1) \cdot k \Bigr) }{k} - \frac{2}{k} \right) \cdot \eps \\
&= \left( \frac{ (\sqrt{2} - 1) \cdot k }{k} - \frac{2}{k} \right) \cdot \eps.
\end{align*}
The first line of this inequality uses the fact that we have no well-positioned
points with color 3, and that the poorly-positioned points can reduce the sum by
at most $\frac{2 \cdot \eps}{k}$. The second line substitutes the bounds that we
have for $x_{1j}$ and $x_{2j}$. The third line uses the fact that $\sqrt{2} - 1$
is a solution of the equation $x = 1 - x - x^2$.

As in the other two cases, this means that we can choose a sufficiently large
constant $k$ such that $\| x - F(X) \|_\infty > (\sqrt{2} - 1) \cdot \eps'$.
\end{enumerate}
\end{enumerate}

Next we observe that the arguments given above all continue to hold if we
substitute a sufficiently precise rational approximation $\sqrt{2}$ in our
displacement vector calculation. This is because all three arguments prove that
some expression converges to $(\sqrt{2} - 1) \cdot \eps > (\sqrt{2} - 1) \cdot
\eps'$, thus we can replace $\sqrt{2}$ with any suitably close rational that
ensures that the expressions converge to  
$(x - 1) \cdot \eps > (\sqrt{2} - 1) \cdot \eps'$ for some $x$.

So far we have shown that there exist three well-positioned points $p_{i_1}$,
$p_{i_2}$, and $p_{i_3}$
that have
three distinct colors. To see that $x$ is contained within a trichromatic
square, it suffices to observe that $\| p_k - p_1 \|_\infty \le 1/2^k$,
which means that all three points must be contained in squares that are adjacent
to the square containing $x$.
\end{proof}

Since $\eps$ can be fixed to be any constant strictly less than $0.5$, we obtain
the following.

\begin{theorem}
\label{thm:2dhardness}
Given a \twoDBrouwer instance, it is
\PPAD-hard to find a point $x \in [0, 1]^2$ such that
\begin{equation*}
\label{eq:2dhardness}
\|x - F(x) \|_\infty < \frac{\sqrt{2} - 1}{2} \approx 0.2071.
\end{equation*}
\end{theorem}

It should be noted that an $\eps$-approximate fixed point can be found in
polynomial time if the function has a suitably small Lipschitz constant, by
trying all points in a grid of width $\eps$. We are
able to obtain a lower bound for constant $\eps$ because our functions have
exponentially large Lipschitz constants.

\paragraph{\bf Improved hardness for \twoDlinFIXP}
Our hardness result for \twoDBrouwer immediately implies the same lower bound for \twoDlinFIXP, 
since the gates in the set $\{c, \badd,
\bsub, \bmul c\}$ can be simulated by the gates in the set $\{c, +, -, * c, \max, \min\}$.
However, we note that a relatively simple modification of our construction can be used to show a tight lower bound for \twoDlinFIXP. In the following theorem we show that it is \PPAD-hard to find a solution for \twoDlinFIXP for all $\eps < 0.5$. This is tight, because $(0.5, 0.5)$ is always a $0.5$-fixed point in any \twoDlinFIXP instance.

\begin{theorem}
\label{thm:linfixp-tight}
\twoDlinFIXP is \PPAD-hard for all $\eps < 0.5$.
\end{theorem}
\begin{proof}
The idea is to use the same same construction as we used for 
Theorem~\ref{thm:2dhardness}, but to set $\eps = 0.5 - \delta$, and then to change the computation of 
$\disp_x$ and $\disp_y$ to multiply the displacement by 100 rather than $\eps$ and then to clip 
the output to the $[0, 1]^2$ square. By multiplying by 100,  we ensure that the output will lie on the boundary of the square whenever one of the colors is missing. Note that since we are considering \twoDlinFIXP, the fact that the intermediate computations used to compute $\disp_x$ and $\disp_y$ will produce values that are not in $[0, 1]$ does not matter, since the gates from \twoDlinFIXP permit this.

More formally, we build the function $F : [0, 1]^2 \rightarrow [0, 1]^2$ such that $F(x, y) = (x', y')$ where
\begin{align*}
x' &= \min(1, \max(0, x + \disp_x \cdot \frac{100}{\eps})), \\
y' &= \min(1, \max(0, y + \disp_y \cdot \frac{100}{\eps})). 
\end{align*}
In this construction, we use $\eps = 100$, and we use
the gates 
$\{c, +, -, * c, \max, \min\}$ 
from \twoDlinFIXP
to simulate 
the bounded gates 
$\{c, \badd, \bsub, \bmul c\}$ where appropriate.
Importantly, we do not use bounded arithmetic when computing $\disp_x$ and $\disp_y$, to ensure that these values are clipped to $[0, 1]$ only at the end.

To prove correctness, we only need to show that the analogue of
Lemma~\ref{lem:ksamples} continues to hold for this new displacement definitions.
It is clear that no solutions can lie in the boundary regions. For the points $x$ that lie inside the boundary, we argue that if $\| x - F(x) \|_\infty \le 0.5 - \delta$, then there are well-positioned samples of all three colors. For the sake of contradiction suppose that this is not the case, which means that at least one color is missing. We have two cases:

\begin{enumerate}
\item First suppose that color 1 is missing. Since there are at most two points
with poorly-positioned coordinates,  we know that we have at least
$k - 2$ points $j$ for which $x_{2j} = 1$ or $x_{3j} = 1$. Hence we have
\begin{equation*}
\disp_y \le \left( \frac{(1 - \sqrt{2} ) (k-2)}{k} + \frac{2}{k} \right) \cdot 100.
\end{equation*}
It is clear that $k$ can be chosen to be suitably large so that $\disp_y \le -1$, implying that $F(x)$ lies on the edge of the $[0, 1]^2$ square. Since $x$ lies inside the boundary region, we have that $\| x - F(x) \|_\infty \ge 0.5 - \eps$.

\item Now suppose that one of colors 2 or 3 is missing. We will consider the
case where color 3 is missing, as the other case is symmetric.
As before, since there are at most two points
with poorly-positioned coordinates,  we know that we have at least
$k - 2$ points $j$ for which $x_{1j} = 1$ or $x_{2j} = 1$.
One of the two following cases applies.

\begin{enumerate}
\item At least $0.1 \cdot k$ well-positioned points satisfy
$x_{2j} = 1$. If this is the case, then we have
\begin{equation*}
\disp_x \ge \left( \frac{0.1 \cdot k}{k} - \frac{2}{k} \right) \cdot
100, \\
\end{equation*}
where we have used the fact that there are no well positioned points with color
3, and the fact that the poorly-positioned points cannot reduce the sum by more
than $\frac{2 }{k}$. 
It is clear that we can choose a suitably large constant $k$ such that $\disp_x \ge 1$, meaning that 
$F(x)$ lies on the edge of the $[0, 1]^2$ square. Since $x$ lies inside the boundary region, we have that $\| x - F(x) \|_\infty \ge 0.5 - \eps$.

\item At least $0.9 \cdot k$ well-positioned points
satisfy $x_{1j} = 1$.
In this case we have
\begin{align*}
\disp_y &\ge \left(\frac{0.9 \cdot k - (\sqrt{2} - 1) \cdot 0.1 \cdot k}{k} - \frac{2}{k}  \right) \cdot 100\\
        &\ge \left(\frac{0.85 \cdot k}{k} - \frac{2}{k} \right) \cdot 100.
\end{align*}
The first line of this inequality uses the fact that we have no well-positioned
points with color 3, and that the poorly-positioned points can reduce the sum by
at most $\frac{2 }{k}$. The second line uses the fact that $0.9 - (\sqrt{2} - 1) \cdot 0.1) \ge 0.85$.

It is clear that we can choose a suitably large constant $k$ such that $\disp_y \ge 1$, and as
in the other two cases, this means that 
we have $\| x - F(x) \|_\infty \ge 0.5 - \eps$. 
\end{enumerate}
\end{enumerate}

\end{proof}


\section{Hardness of \twoDBrouwer with a constant width circuit}
\label{sec:thincircuit}

In our reduction from \twoDBrouwer to tree polymatrix games, the number of
actions in the game will be determined by the width of the circuit.
This means that the hardness proof from the previous section is not a sufficient
starting point,
because it
produces \twoDBrouwer instances that have circuits with high width. In
particular, the circuits will extract $2n$ bits from the two inputs, which means
that the circuits will have width at least $2n$. 

Since we desire a constant number of actions in our tree polymatrix game, we
need to build a hardness proof for \twoDBrouwer that produces a circuit with
constant width. In this section we do exactly that, by reimplementing the
reduction from the previous section using gadgets that keep the width small.

\myparagraph{\bf Bit packing.}
We adopt an idea of \cite{EGG}, to store many
bits in a single arithmetic value using a \emph{packed} representation. Given
bits $b_1, b_2, \dots, b_k \in \{0, 1\}$, the packed representation of these
bits is the value 
$\packed(b_1, b_2, \dots, b_k) := \sum_{i = 1}^k b_i/2^i$.
We will show that the reduction from
the previous section can be performed while keeping all Boolean values in a
single variable that uses packed representation.

\myparagraph{\bf Working with packed variables.}
We build SLPs that work with this packed representation, two of which are shown
below. 
\begin{center}
\scalebox{1}{
\begin{minipage}[c]{0.5\textwidth}
\begin{algorithm2e}[H]
\caption{$\mathtt{FirstBit(x, b)}$ \\+0 variables}
\label{slp:firstbit}
\DontPrintSemicolon
\tcp{Extract the first bit of x into \texttt{b}}
\texttt{b $\leftarrow$ 0.5}\;
\texttt{b $\leftarrow$ x \bsuba b}\;
\texttt{b $\leftarrow$ b \bmula L}\;
\;
\tcp{Remove the first bit of \texttt{x}}
\texttt{b $\leftarrow$ b \bmula 0.5}\;
\texttt{x $\leftarrow$ x \bsuba b}\;
\texttt{x $\leftarrow$ x \bmula 2}\;
\texttt{b $\leftarrow$ b \bmula 2}\;

\end{algorithm2e}
\end{minipage}
}%
\scalebox{1}{
\begin{minipage}[c]{0.4\textwidth}
\begin{algorithm2e}[H]
\caption{$\mathtt{Clear(I, x)}$ \\ +2 variables}
\label{slp:clear}
\DontPrintSemicolon
\texttt{x' $\leftarrow$ x \bmula 1} \;
\For{\texttt{i in  $\{1, 2, \dots, k\}$}}{
    \texttt{b $\leftarrow$ 0} \;
    \texttt{FirstBit(x', b)} \;
    \If{\texttt{i $\in$ I}}{
       \texttt{b $\leftarrow$ b \bmula $\frac{1}{2^i}$ }\;
       \texttt{x $\leftarrow$ x \bsuba b }\;
    }
}

\end{algorithm2e}
\end{minipage}
}
\end{center}
The \texttt{FirstBit} SLP (SLP~\ref{slp:firstbit}) 
combines the ideas from SLPs~\ref{slp:extract} and~\ref{slp:extract-multi} to
extract the first bit from a value $x \in [0, 1]$. Repeatedly applying this SLP
allows us to read out each bit of a value in sequence. 
%
%
The \texttt{Clear} SLP (SLP~\ref{slp:clear}) uses this to set some bits of a packed variable
to zero. It takes as input a set of indices $I$, and a packed variable $x =
\packed(b_1, b_2, \dots, b_k)$. At the end of the SLP we have $x = \packed(b'_1,
b'_2, \dots, b'_k)$ where $b'_i = 0$ whenever $i \in I$, and $b'_i = b_i$
otherwise.

It first copies $x$ to a fresh variable $x'$. The bits of $x'$
are then read-out using \texttt{FirstBit}. Whenever a bit $b_i$
with $i \in I$ is decoded from $x'$, we subtract $b_i/2^{i}$ from $x$. If $b_i =
1$, then this sets the corresponding bit of $x$ to zero, and if $b_i = 0$, then
this leaves $x$ unchanged.

We want to minimize the width of the circuit that we produce, so we keep track of
the number of \emph{extra} variables used by our SLPs.
For \texttt{FirstBit},
this is zero, while for
\texttt{Clear} this is two, since that SLP uses the fresh variables $x'$ and $b$.



\myparagraph{Packing and unpacking bits.}
We implement two SLPs that manipulated packed variables.
The \texttt{Pack(x,y,S)} operation allows us to extract bits from $y \in [0,1]$, 
and store them in $x$, while the 
\texttt{Unpack(x,y,S)} operation allows us to extract bits from $x$ to create
a value $y \in [0, 1]$. This is formally specified in the following lemma.


\begin{lemma}
\label{lem:packunpack}
Suppose that we are given $\mathtt{x} = \packed(b_1, b_2, \dots,
b_k)$, a variable $\mathtt{y} \in [0, 1]$, and a sequence of indices $S = \langle s_1,
s_2, \dots, s_j \rangle.$ Let $y_j$ denote the $j$th bit of $y$. The following SLPs can be implemented using at most
two extra variables.
\begin{itemize}
\item \texttt{Pack(x, y, S)} modifies $\mathtt{x}$ so that $\mathtt{x} =
\packed(b'_1, b'_2, \dots, b'_k)$ where $b'_{i} = y_{j}$ whenever there exists
an index $s_j \in S$ with $s_j = i$, and $b'_{i} = b_i$ otherwise.
\item \texttt{Unpack(x, y, S)} modifies \texttt{y} so that $\mathtt{y} = \mathtt{y} \badd \sum_{i = 1}^j b_{s_i}/2^i$
\end{itemize}
\end{lemma}

\begin{proof}

We construct SLPs for both of the operations.



\myparagraph{\bf Packing bits.}
SLP~\ref{slp:pack} implements the pack operation. It begins by clearing the bits
referenced by the sequence~$S$. It then copies \texttt{y} to \texttt{y'}, and
destructively extracts the first $j$ bits of \texttt{y'}. These bits are then
stored at the correct index in \texttt{x} by the final line of the \texttt{for}
loop. In total, this SLP uses two additional variables \texttt{y'} and
\texttt{b}. Two extra variables are used by \texttt{Clear}, but these stop being
live after the first line, before \texttt{y'} and \texttt{b} become live. 

\myparagraph{\bf Unpacking bits.} 
SLP~\ref{slp:unpack} implements the unpacking operation. It first copies
\texttt{x} to \texttt{x'}, and then destructively extracts the first $k$ bits of
\texttt{x'}. Whenever a bit referred to by $S$ is extracted from \texttt{x'}, it
is first multiplied by $\frac{1}{2^{s_j}}$, which puts it at the correct
position, and is then added to \texttt{y}. This SLP uses the two additional
variables \texttt{x'} and \texttt{b}.
\end{proof}

\begin{center}
\begin{minipage}[c]{0.4\textwidth}
\begin{algorithm2e}[H]
\caption{$\mathtt{Pack(x, y, S)}$ \quad +2 variables}
\label{slp:pack}
\DontPrintSemicolon
\texttt{Clear(S, x)} \;
\texttt{y' $\leftarrow$ y \bmula 1} \;
\For{\texttt{i in  $\{1, 2, \dots, j\}$}}{
    \texttt{b $\leftarrow$ 0} \;
    \texttt{FirstBit(y',b)} \;
    \texttt{x $\leftarrow$ b \bmula $\frac{1}{2^{s_i}}$} \;
}
\end{algorithm2e}
\end{minipage}\hskip 1cm%
\begin{minipage}[c]{0.4\textwidth}
\begin{algorithm2e}[H]
\caption{$\mathtt{Unpack(x, y, S)}$ \quad +2 variables}
\label{slp:unpack}
\DontPrintSemicolon
\texttt{x' $\leftarrow$ x \bmula 1} \;
\For{\texttt{i in  $\{1, 2, \dots, k\}$}}{
    \texttt{b $\leftarrow$ 0} \;
    \texttt{FirstBit(x', b)} \;
    \If{\texttt{i = s$_j$ for some j}}{
    $\texttt{b $\leftarrow$ b \bmula $\frac{1}{2^{s_j}}$}$ \;
    $\texttt{y $\leftarrow$ y \badda b}$ \;
    }
}
\end{algorithm2e}
\end{minipage}
\end{center}


\myparagraph{\bf Simulating a Boolean operations.} 
As described in the previous section, the reduction only needs to simulate or-
and not-gates. 
Given $\mathtt{x} = \packed(b_1, b_2, \dots, b_k)$, and three indices $i_1, i_2, i_3$, we
implement two SLPs, which both modify $x$ so that
$\mathtt{x} = \packed(b'_1, b'_2, \dots, b'_k)$.
SLP~\ref{slp:or} implements $\mathtt{Or(x, i_1, i_2, i_3)}$, which ensures that
$b'_{i_3} = b_{i_1} \lor
b_{i_2}$, and $b'_i = b_i$ for $i \ne i_3$.
SLP~\ref{slp:not} implements $\mathtt{Not(x, i_1, i_2)}$, which ensures that
$b'_{i_2} = \lnot b_{i_1}$, and $b'_i = b_i$ for $i \ne i_2$.

\begin{center}
\begin{minipage}[c]{0.4\textwidth}
\begin{algorithm2e}[H]
\caption{$\mathtt{Or(x, i_1, i_2, i_3)}$ +3 variables}
\label{slp:or}
\DontPrintSemicolon
\texttt{a $\leftarrow$ 0} \;
\texttt{Unpack(x,a,$\langle \mathtt{i_1} \rangle$)} \;
\texttt{Unpack(x,a,$\langle \mathtt{i_2} \rangle$)} \;
\texttt{Pack(x,a,$\langle \mathtt{i_3} \rangle$)} \;
\end{algorithm2e}
\end{minipage}\hskip 1cm%
\begin{minipage}[c]{0.4\textwidth}
\begin{algorithm2e}[H]
\caption{$\mathtt{Not(x, i_1, i_2)}$ +3 variables}
\label{slp:not}
\DontPrintSemicolon
\texttt{a $\leftarrow$ 0} \;
\texttt{Unpack(x,a,$\langle \mathtt{i_1} \rangle$)} \;
\texttt{b $\leftarrow$ 1} \;
\texttt{a $\leftarrow$ b \bsuba a} \;
\texttt{Pack(x,a,$\langle \mathtt{i_2} \rangle$)} \;
\end{algorithm2e}
\end{minipage}
\end{center}

These two SLPs simply unpack the input bits, perform the operation, and then
pack the result into the output bit. The \texttt{Or} SLP uses the
\texttt{Unpack} operation to set $\mathtt{a} = b_{i_1} \badd b_{i_2}$.
Both SLPs use three extra variables: the fresh variable \texttt{a} is live
throughout, and the pack and unpack operations use two extra variables. The
variable \texttt{b} in the \texttt{Not} SLP is not live concurrently with a pack
or unpack, and so does not increase the number of live variables.
These two SLPs can be used to simulate a Boolean circuit using at most three
extra variables.

\begin{lemma}
\label{lem:circuit}
Let $C$ be a Boolean circuit with $n$ inputs and $k$ gates. Suppose that $x =
\packed(b_1, \dots, b_n)$, gives values for the inputs of the circuit. There is
an SLP \texttt{Simulate(C, x)} that uses three extra variables, and modifies $x$
so that $x = \packed(b_1, \dots, b_n, b_{n+1}, \dots, b_{n+k})$, where $b_{n+i}$
is the output of gate $i$ of the circuit.
\end{lemma}
\begin{proof}
\myparagraph{\bf Simulating a Boolean circuit.} 
Let $\langle g_{n+1}, g_{n+2}, \dots, g_{n+k} \rangle$ be the gates of the
circuit, and suppose, without loss of generality, that the gates have been
topologically ordered. SLP~\ref{slp:circuit} will simulate the circuit $C$.


\begin{center}
\scalebox{1}{
\begin{minipage}[c]{0.8\textwidth}
\begin{algorithm2e}[H]
\caption{$\mathtt{Simulate(C, x)}$ \\ +3 variables}
\label{slp:circuit}
\DontPrintSemicolon
\For{\texttt{i in  $\{n+1, n+2, \dots, n+k\}$}}{
    \If{\texttt{$g_{i}$ = $g_{j_1} \lor g_{j_2}$}}{
        \texttt{Or(x,i,j$_1$,j$_2$)}
    }
    \If{\texttt{$g_{i}$ = $\lnot g_{j}$}}{
        \texttt{Not(x,i,j)}
    }
}
\end{algorithm2e}
\end{minipage}
}
\end{center}
Assuming that the first $n$ bits of $x$ already contain the packed inputs of the
circuit, SLP~\ref{slp:circuit} implements the operation  $\mathtt{Simulate(C,
x)}$ that computes the output of each gate. This simply iterates through and
simulates each gate. The SLP introduce no new variables, and so it uses three
additional live variables in total, which come from the \texttt{Or} and
\texttt{Not} operations.
\end{proof}

\myparagraph{\bf Implementing the reduction.}
Finally, we can show that the circuit built in \cref{thm:2dhardness} can
be implemented by an SLP that uses at most 8 variables. This SLP 
cycles through each sampled point in turn, computes the $x$ and $y$ displacements by
simulating the Boolean circuit, and then adds the result to the output.




\begin{theorem}
\label{thm:2dhardness-lw}
Given a \twoDBrouwer instance, it is
\PPAD-hard to find a point $x \in [0, 1]^2$ with
$\| x - F(x) \|_\infty < \frac{\sqrt{2} - 1}{2}$ 
even for a synchronous circuit of width eight.
\end{theorem}
\begin{proof}

\myparagraph{\bf Dealing with the output.}
Recall that our Boolean circuit will output three bits, and that these bits
determine which displacement vector is added to the output of the arithmetic
circuit.  We now build an SLP that does this conversion. It implements 
$\mathtt{AddVector(x, i, \outp_x, \outp_y, k, d_x, d_y)}$, where $x = \packed(b_1, b_2,
\dots, b_n)$, $i \le n$ is an index, $\outp_x$ and $\outp_y$ are variables, $k$
is an integer, and
$d_x, d_y \in [-1, 1]$. After this procedure, we should have $\outp_x = \outp_x
+ d_x \cdot b_i / k$, and $\outp_y = \outp_y + d_y \cdot b_i / k$.
SLP~\ref{slp:addvector} does this operation. It uses three extra variables in
total: the fresh variable \texttt{a} is live throughout, and the two
unpack operations use two extra variables.

\begin{center}
\scalebox{1}{
\begin{minipage}[c]{0.8\textwidth}
\begin{algorithm2e}[H]
\caption{$\mathtt{AddVector(x, i, \outp_x, \outp_y, k, d_x, d_y)}$ \\  +3 variables}
\label{slp:addvector}
\DontPrintSemicolon
\tcp{Add $\mathtt{d_x \cdot b_i}$ to $\outp_x$}
\texttt{a $\leftarrow$ 0} \;
\texttt{Unpack(x,a,$\langle i \rangle$)} \;
\texttt{a $\leftarrow$ $\mathtt{|d_x|/k}$ \bmula a} \;
\texttt{$\outp_x$ $\leftarrow$ $\outp_x$ \badda a} \tcp*{Use $\bsuba$ if d$_x$ $<$ 0} \;

\tcp{Add $\mathtt{d_y \cdot b_i}$ to $\outp_y$}
\texttt{a $\leftarrow$ 0} \;
\texttt{Unpack(x,a,$\langle i \rangle$)} \;
\texttt{a $\leftarrow$ $\mathtt{|d_y|/k}$ \bmula a} \;
\texttt{$\outp_y$ $\leftarrow$ $\outp_y$ \badda a} \tcp*{Use $\bsuba$ if d$_y$ $<$ 0} \;
\end{algorithm2e}
\end{minipage}
}
\end{center}

\myparagraph{\bf Implementing the reduction.} Finally, we can implement the
reduction from \discreteBrouwer to \twoDBrouwer. We will assume that we have
been given a Boolean circuit $C$ that takes $2n$ inputs, where the first $n$
input bits correspond to the $x$ coordinate, and the second $n$ input bits
correspond to the $y$ coordinate. Recall that we have required that $C$ gives
its output as a one-hot vector. We assume that the three output bits of $C$ are
indexed $n+k-2$, $n+k-1$, and $n+k$, corresponding to colors $1$, $2$, and $3$,
respectively.

\begin{center}
\scalebox{1}{
\begin{minipage}[c]{0.8\textwidth}
\begin{algorithm2e}[H]
\caption{$\mathtt{Reduction(\inp_x, \inp_y, \outp_x, \outp_y)}$ \\ +4 variables}
\label{slp:reduction}
\DontPrintSemicolon
\texttt{$\outp_x$ $\leftarrow$ $\inp_x$} \;
\texttt{$\outp_y$ $\leftarrow$ $\inp_y$} \;
\For{\texttt{i in  $\{1, 2, \dots, k\}$}}{
    \texttt{$\inp_x$ $\leftarrow$ $\inp_x$ \badda $1/((k+1) \cdot 2^{n+1})$} \;
    \texttt{$\inp_y$ $\leftarrow$ $\inp_y$ \badda $1/((k+1) \cdot 2^{n+1})$} \;
    \texttt{x $\leftarrow$ 0} \;
    \texttt{Pack(x, $\inp_x$, $\mathtt{\langle 1, 2, \dots, n \rangle}$)} \;
    \texttt{Pack(x, $\inp_y$, $\mathtt{\langle n+1, n+2, \dots, 2n \rangle}$)} \;
    \texttt{Simulate(C, x)} \;
    \texttt{AddVector(x, n+k-2, $\mathtt{\outp_x}$, $\mathtt{\outp_y}$, k, \phantom{-}0, 1)}\;
    \texttt{AddVector(x, n+k-1, $\mathtt{\outp_x}$, $\mathtt{\outp_y}$, k, \phantom{-}1, 1-$\sqrt{\mathtt{2}}$)}\;
    \texttt{AddVector(x, n+k\phantom{-1}, $\mathtt{\outp_x}$,
$\mathtt{\outp_y}$, k, -1, 1-$\sqrt{\mathtt{2}}$)}\;
}
\end{algorithm2e}
\end{minipage}
}
\end{center}
%
SLP~\ref{slp:reduction} implements the reduction. The variables $\inp_x$ and
$\inp_y$ hold the inputs to the circuit, while the variables $\outp_x$ and
$\outp_y$ are the outputs. The SLP first copies the inputs to the outputs, and
then modifies the outputs using the displacement vectors. Each iteration of the
\texttt{for} loop computes the computes the displacement contributed by the
point $p_i$ (defined in the previous section). This involves decoding the first
$n$ bits of both $\inp_x$ and $\inp_y$, which can be done via the pack
operation, simulating the circuit on the resulting
bits, and then adding the correct displacement vectors to $\outp_x$ and
$\outp_y$.

The correctness of this SLP follows from our correctness proof for
\cref{thm:2dhardness}, since all we have done in this section is
reimplement while using a small number of live variables.
In particular, we note that the implementation that we give here will not be
broken by poorly positioned points. In any iteration of the for loop in which
one of $\inp_x$ or $\inp_y$ are poorly positioned, the $\texttt{Pack}$ operation may
fail to extract pure bits, and if that occurs, then the resulting value stored in $\texttt{x}$ will be
nonsensical. This will, in turn, cause the operations used by
$\texttt{Simulate}$ to
give incorrect outputs, and then the input to $\texttt{AddVector}$ will
likewise be incorrect. Indeed, $\texttt{x}$ may never contain a valid packed
encoding at any step of the procedure.

However, note that the variable \texttt{a} used by $\texttt{AddVector}$ when it
is applied to \texttt{x} will still lie in the range $[0, 1]$, even when the
input is not a valid packed encoding. This is because all of our operations are
ultimately implemented by operations that bound the result to lie in $[0, 1]$.
Hence, in the iterations that consider poorly positioned points, the for loop
will treat the output of $C$ as being some aribtary vector in $[0, 1]^3$. This
is exactly the same assumption that was made in the proof of
\cref{thm:2dhardness}, so the correctness of that proof is unaffected.

In total, this SLP uses four extra variables. All of the macros use at most
three extra variables, and the fresh variable \texttt{x} during these macros.
Since 
$\inp_x$ $\inp_y$, $\outp_x$ and
$\outp_y$ are all live throughout as well, this gives us 8 live variables in
total. 
\end{proof}

\section{Hardness for tree polymatrix games}

Now we show that finding a Nash equilibrium of a tree polymatrix
game is \PPAD-hard. We reduce from the low-width \twoDBrouwer
problem, whose hardness was shown in Theorem~\ref{thm:2dhardness-lw}. Throughout
this section, we suppose that we have a \twoDBrouwer instance defined by a synchronous
arithmetic circuit $F$ of width eight and depth $n$. The gates of this circuit will be 
indexed as $g_{i,j}$ where $1 \le i \le 8$ and $1 \le j \le n$, meaning that
$g_{i,j}$ is the $i$th gate on level $j$.

\myparagraph{\bf Modifying the circuit.}
The first step of the reduction is to modify the circuit.
First, we modify the circuit so that all gates operate on values in
$[0, 0.1]$, rather than $[0, 1]$. We introduce the operators $\badd_{0.1}$,
$\bsub_{0.1}$, and $\bmul_{0.1}$, which bound their outputs to be in
$[0, 0.1]$. 
The following lemma, states that we
can rewrite our circuit using these new gates. 
The
transformation simply divides all $c$-gates in the circuit by ten.

\begin{lemma}
\label{lem:dividebyten}
Given an arithmetic circuit $F : [0, 1]^2 \rightarrow [0, 1]^2$ that uses gates
from
$\{c, \badd, \bsub, \bmul\}$, 
we can construct a circuit $F' : [0, 0.1]^2
\rightarrow [0, 0.1]^2$ that uses the gates from
$\{c, \badd_{0.1}, \bsub_{0.1}, \bmul_{0.1}\}$, so that $F(x, y)
= (x, y)$ if and only if $F'(x/10, y/10) = (x/10, y/10)$.
\end{lemma}
\begin{proof}
The circuit $F'$ consists of gates $g'_{i,j}$ for 
each $1 \le i \le 8$ and $1 \le j \le n$.
\begin{itemize}
\item If $g_{i,j} = c$, then $g'_{i,j} = c/10$.
\item If $g_{i,j} = g_{a,b} \badd g_{x,y}$, then 
$g'_{i,j} = g'_{a,b} \; \badd_{0.1} \; g'_{x,y}$.
\item If $g_{i,j} = g_{a,b} \bsub g_{x,y}$, then 
$g'_{i,j} = g'_{a,b} \; \bsub_{0.1} \; g'_{x,y}$.
\item If $g_{i,j} = g_{a,b} \bmul c$, then 
$g'_{i,j} = g'_{a,b} \; \bmul_{0.1} \; c$.
\end{itemize}
Let $(x, y) \in [0, 1]^2$. It is not difficult to show by induction, that if we
compute $F(x, y)$ and $F'(x/10, y/10)$, then $g'_{i,j} = g_{i,j}/10$ for all $i$
and $j$. Hence, $F'(x/10, y/10) = (x/10,
y/10)$ if and only if $F(x, y) = (x, y)$.
\end{proof}

\begin{figure}
\begin{center}
\scalebox{0.7}{ 
\tikzstyle{state}=[minimum size=1.1cm,inner sep=0cm,draw,circle,node distance=2cm]

\begin{tikzpicture}

	\node [state]             	(g11){$g_{1,1}$};
	\node [state,right of=g11]  (g12){$g_{2,1}$};

	\node [right of=g12, node distance=2cm]  (dots1){$\cdots$};
	\node [state,right of=g12, node distance=4cm]  (g17){$g_{7,1}$};
	\node [state,right of=g17]  (g18){$g_{8,1}$};
	\node [state,right of=g18]  (g19){$g_{9,1}$};
	\node [state,right of=g19]  (g10){$g_{10,1}$};

	\node [state,above of=g11] 	(g21){$g_{1,2}$};
	\node [state,above of=g12]  (g22){$g_{2,2}$};
	\node [right of=g22, node distance=2cm]      (dots2){$\cdots$};
	\node [state,above of=g17]  (g27){$g_{7,2}$};
	\node [state,above of=g18]  (g28){$g_{8,2}$};
	\node [state,above of=g19]  (g29){$g_{9,2}$};
	\node [state,above of=g10]  (g20){$g_{10,2}$};

	\node [above of=g21, node distance=2cm]  (g31){};
	\node [above of=g22, node distance=2cm]  (g32){};
	\node [above of=g27, node distance=2cm]  (g37){};
	\node [above of=g28, node distance=2cm]  (g38){};
	\node [above of=g29, node distance=2cm]  (g39){};
	\node [above of=g20, node distance=2cm]  (g30){};

	\node [right of=g32, node distance=2cm]  (dots3){};

	\node [above of=g31, node distance=2cm]  (g41){};
	\node [above of=g32, node distance=2cm]  (g42){};
	\node [above of=g37, node distance=2cm]  (g47){};
	\node [above of=g38, node distance=2cm]  (g48){};
	\node [above of=g39, node distance=2cm]  (g49){};
	\node [above of=g30, node distance=2cm]  (g40){};

	\node at ($0.5*(g31) + 0.5*(g41)$) {$\vdots$};
	\node at ($0.5*(g32) + 0.5*(g42)$) {$\vdots$};
	\node at ($0.5*(g37) + 0.5*(g47)$) {$\vdots$};
	\node at ($0.5*(g38) + 0.5*(g48)$) {$\vdots$};
	\node at ($0.5*(g39) + 0.5*(g49)$) {$\vdots$};
	\node at ($0.5*(g30) + 0.5*(g40)$) {$\vdots$};

	\node [state,above of=g41]  (gn1){$g_{1,n}$};
	\node [state,above of=g42]  (gn2){$g_{2,n}$};
	\node [state,above of=g47]  (gn7){$g_{7,n}$};
	\node [state,above of=g48]  (gn8){$g_{8,n}$};
	\node [state,above of=g49]  (gn9){$g_{9,n}$};
	\node [state,above of=g40]  (gn0){$g_{10,n}$};

	\node [right of=gn2, node distance=2cm]      (dotsn){$\cdots$};

	\path[->, thick] (gn7.north) edge[out=90, in=90] (gn9.north);
	\path[->, thick] (gn8.north) edge[out=90, in=90] (gn0.north);

	\path[->, thick] (g19.south) edge[out=-90, in=-90] (g17.south);
	\path[->, thick] (g10.south) edge[out=-90, in=-90] (g18.south);

	\path[->, thick] (gn9.south) edge (g49.north);
	\path[->, thick] (gn0.south) edge (g40.north);

	\path[->, thick] (g39.south) edge (g29.north);
	\path[->, thick] (g30.south) edge (g20.north);

	\path[->, thick] (g29.south) edge (g19.north);
	\path[->, thick] (g20.south) edge (g10.north);


\path[->, dashed]
    (g21) edge (g32)
    (g21) edge (g31)
    (g11) edge (g21)
    (g11) edge (g22)
    (g12) edge (g21)
    (g12) edge (g27)
    (g17) edge (g28)
    (g18) edge (g28)
    (g18) edge (g27)
    (g22) edge (dots3)
    (g27) edge (g37)
    (g28) edge (g38)

    (g41) edge (gn1)
    (g42) edge (gn1)
    (g42) edge (gn2)
    (g47) edge (gn2)
    (g47) edge (gn7)
    (g48) edge (gn8)
    ;

\end{tikzpicture}

}
\caption{Extra equalities to introduce feedback of $g_{7,n}$ and $g_{8,n}$ to
$g_{7,1}$ and $g_{8,1}$ respectively.}
\label{fig:loopback}
\end{center}
\end{figure}

Next we modify the structure of the circuit by connecting the
two outputs of the circuit to its two inputs. 
Suppose, without loss of generality, that
$g_{7,1}$ and $g_{8,1}$ are the inputs and that $g_{7,n}$
and $g_{8,n}$ are outputs. Note that the equality $x = y$ can be
implemented using the gate $x = y \bmul_{0.1} 1$. We add the following extra
equalities, which are shown in \cref{fig:loopback}.
\begin{itemize}
\item We add gates $g_{9,n-1} = g_{7,n}$ and $g_{10, n-1} = g_{8,n}$.
\item 
For each $j$ in the range $2 \le j < n-1$, we add $g_{9,j} = g_{9,j+1}$
and $g_{10,j} = g_{10,j+1}$.
\item 
We modify $g_{7,1}$ so that $g_{7,1} = g_{9,2}$, and we modify $g_{8,1}$
so that $g_{8,1} = g_{10, 2}$.
\end{itemize}
Note that these gates are backwards: they copy values from higher levels in the
circuit to lower levels, and so the result is not a circuit, but a system of
constraints defined by gates, with  
some structural properties. Firstly,
each gate $g_{i,j}$ is only involved in constraints with gates of the form
$g_{i',j+1}$ and $g_{i',j-1}$.
Secondly, finding values for the gates that satisfy all of the constraints is
\PPAD-hard, since by construction such values would yield a fixed point of $F$.

\myparagraph{\bf The polymatrix game.}
The polymatrix game will contain three types of players.
\begin{itemize}
\item For each $i= 1,\ldots,n$, we have a \emph{variable} player $v_i$.
\item For each $i= 1,\ldots,n-1$, we have a \emph{constraint}
player $c_i$, who is connected to $v_i$ and $v_{i+1}$.
\item For each $i=1,\ldots,2n-1$, we have a \emph{mix} player
$m_i$. If $i$ is even, then $m_i$ is connected to~$c_{i/2}$. If $i$ is odd, then
$m_i$ is connected to $v_{(i+1)/2}$.
\end{itemize}
The structure of this game is shown in \cref{fig:structure}.
Each player has twenty actions, which are divided into ten pairs,
$x_i$ and $\bx_i$ for $i = 1,\ldots,10$. 
\begin{figure}
\begin{center}
\scalebox{0.7}{ 
\begin{tikzpicture}
\node [state]             (v1) {$v_1$};
\node [state,below of=v1] (m1) {$m_1$};

\node [state,right of=v1] (c1) {$c_1$};
\node [state,below of=c1] (m2) {$m_2$};

\node [state,right of=c1] (v2) {$v_2$};
\node [state,below of=v2] (m3) {$m_3$};

\node [state,right of=v2] (c2) {$c_2$};
\node [state,below of=c2] (m4) {$m_4$};

\node [state,right of=c2] (v3) {$v_3$};
\node [state,below of=v3] (m5) {$m_5$};

\node [right of=v3,node distance=2cm,minimum size=2cm] (mid) {$\dots$};

\node [state,right of=v3,node distance=4cm] (vn) {$v_n$};
\node [state,below of=vn] (mn) {$m_{2n-1}$};

\path 
    (v1) edge (m1)
    (c1) edge (m2)
    (v2) edge (m3)
    (c2) edge (m4)
    (v3) edge (m5)
    (vn) edge (mn)

    (v1) edge (c1)
    (c1) edge (v2)
    (v2) edge (c2)
    (c2) edge (v3)

    (v3) edge (mid)
    (mid) edge (vn)
    ;

\end{tikzpicture}
}
\end{center}
\caption{The structure of the polymatrix game.}
\label{fig:structure}
\end{figure}

\myparagraph{\bf Forcing mixing.}
The role of the mix players is to force the variable and constraint players to
play specific mixed strategies: for every 
variable or constraint
player $j$, we want $s_j(x_i) + s_j(\bx_i) = 0.1$ for all $i$,
which means that the same amount of probability is assigned to each pair of
actions.
To force this, each mix player plays a high-stakes hide-and-seek against their
opponent, which is shown in \cref{fig:hide-and-seek}. This zero-sum game is defined by a $20 \times 20$
matrix $Z$ and a constant $M$. The payoff $Z_{ij}$ is defined as follows.
If $i \in \{x_a, \bx_a\}$ and $j \in \{x_a, \bx_a\}$ for some $a$, then
$Z_{ij} = M$.
Otherwise, $Z_{ij} = 0$.
For each $i$ the player $m_i$ plays against player $j$, which is either a
constraint player $c_{i'}$ or a variable player $v_{i'}$. We define the payoff
matrix $A^{m_i, j} = Z$ and $G^{j, m_i} = -Z$.
\begin{figure}
\begin{center}
\scalebox{0.5}{
\pgfmathsetmacro{\sc}{0.35}
\pgfmathsetmacro{\ll}{-0.9}

\pgfmathsetmacro{\os}{0.5}

\begin{tikzpicture}

\matrix (m) [matrix of nodes, 
             	anchor=south west,
			 	ampersand replacement=\&,
				nodes={draw, minimum size=2cm},
				column sep={2cm,between origins},
                row sep={2cm,between origins},
             	nodes in empty cells,
             	inner sep=0pt,
				row 5/.append style={nodes={draw=none}},
				column 5/.append style={nodes={draw=none}},
		]
{ 
	\& \& \& \& \& \& \\
	\& \& \& \& \& \& \\
	\& \& \& \& \& \& \\
	\& \& \& \& \& \& \\ 
	\& \& \& \& \& \& \\ 
	\& \& \& \& \& \& \\ 
	\& \& \& \& \& \& \\ 
};

\payoffs{1}{1}{\Large $M$}{\Large $-M$}
\payoffs{1}{2}{\Large $M$}{\Large $-M$}
\payoffs{2}{1}{\Large $M$}{\Large $-M$}
\payoffs{2}{2}{\Large $M$}{\Large $-M$}
\payoffs{1}{3}{\Large $0$}{\Large $0$}
\payoffs{1}{4}{\Large $0$}{\Large $0$}
\payoffs{2}{3}{\Large $0$}{\Large $0$}
\payoffs{2}{4}{\Large $0$}{\Large $0$}
\payoffs{3}{1}{\Large $0$}{\Large $0$}
\payoffs{3}{2}{\Large $0$}{\Large $0$}
\payoffs{4}{1}{\Large $0$}{\Large $0$}
\payoffs{4}{2}{\Large $0$}{\Large $0$}
\payoffs{3}{3}{\Large $M$}{\Large $-M$}
\payoffs{3}{4}{\Large $M$}{\Large $-M$}
\payoffs{4}{3}{\Large $M$}{\Large $-M$}
\payoffs{4}{4}{\Large $M$}{\Large $-M$}

\rowstrategy{1}{$\bar{x}_1$}
\rowstrategy{2}{$x_1$}
\rowstrategy{3}{$\bar{x}_2$}
\rowstrategy{4}{$x_2$}

\colstrategy{1}{$\bar{x}_1$}
\colstrategy{2}{$x_1$}
\colstrategy{3}{$\bar{x}_2$}
\colstrategy{4}{$x_2$}

\players{$m_i$}{$c_{i/2}$}

\rowstrategy{5}{$\vdots$}
\colstrategy{5}{$\cdots$}
\node at (m-5-5) {\LARGE $\ddots$};

\payoffs{1}{6}{\Large $0$}{\Large $0$}
\payoffs{1}{7}{\Large $0$}{\Large $0$}
\payoffs{2}{6}{\Large $0$}{\Large $0$}
\payoffs{2}{7}{\Large $0$}{\Large $0$}
\payoffs{3}{6}{\Large $0$}{\Large $0$}
\payoffs{3}{7}{\Large $0$}{\Large $0$}
\payoffs{4}{6}{\Large $0$}{\Large $0$}
\payoffs{4}{7}{\Large $0$}{\Large $0$}

\payoffs{6}{1}{\Large $0$}{\Large $0$}
\payoffs{7}{1}{\Large $0$}{\Large $0$}
\payoffs{6}{2}{\Large $0$}{\Large $0$}
\payoffs{7}{2}{\Large $0$}{\Large $0$}
\payoffs{6}{3}{\Large $0$}{\Large $0$}
\payoffs{7}{3}{\Large $0$}{\Large $0$}
\payoffs{6}{4}{\Large $0$}{\Large $0$}
\payoffs{7}{4}{\Large $0$}{\Large $0$}

\payoffs{6}{6}{\Large $M$}{\Large $-M$}
\payoffs{6}{7}{\Large $M$}{\Large $-M$}
\payoffs{7}{6}{\Large $M$}{\Large $-M$}
\payoffs{7}{7}{\Large $M$}{\Large $-M$}

\rowstrategy{6}{$\bar{x}_{20}$}
\rowstrategy{7}{$x_{20}$}

\colstrategy{6}{$\bar{x}_{20}$}
\colstrategy{7}{$x_{20}$}

\end{tikzpicture}

}
\caption{The hide and seek game that forces $c_{j/2}$ to play an appropriate
mixed strategy. The same game is
used to force $v_{(j-1)/2}$ mixes appropriately.}
\label{fig:hide-and-seek}
\end{center}
\end{figure}
%
%
The following lemma, shows that if $M$ is
suitably large, then the variable and constraint players must allocate
probability $0.1$ to each of the ten action pairs.
\begin{lemma}
\label{lem:mixing}
Suppose that all payoffs in the games between variable and constraint players
use payoffs in the range $[-P, P]$. If $M > 40 \cdot P$ then in every mixed Nash
equilibrium $\sbf$, the action $s_j$ of every variable and constraint player
$j$ satisfies $s_j(x_i) + s_j(\bx_i) = 0.1$ for all $i$.
\end{lemma}
\begin{proof}
For the sake of contradiction, suppose that there is a Nash equilibrium $\sbf$
in which there is some variable or constraint player $j$ that fails to satisfy
this equality. Let $I$ be the subset of indices that maximize the
expression $s_j(x_i) + s_j(\bx_i)$, i.e., $I$ contains the pairs that player $j$
plays with highest probability. Note that since player $j$ does not play all
pairs uniformly, $I$ does not contain every index, so let $J$ be the non-empty
set of indices not in $I$.

Let $m_k$ be the mix player who plays against player $j$.
By construction, the actions $x_i$ and $\bx_i$ have payoff $\left(s_j(x_i) +
s_j(\bx_i)\right) \cdot M$ for $m_k$. Since $\sbf$ is a Nash equilibrium, $m_k$ may only place
probability on actions that are best responses, which means that he may
only place probability on the actions $x_i$ and $\bx_i$ when $i \in I$.

Let $i$ be an index that maximizes $s_{m_k}(x_i) + s_{m_k}(\bx_i)$ for player
$m_k$. By the above argument, we have $i \in I$. The actions $x_i$ and
$\bx_i$ for player $j$ give payoff at most

\begin{align*}
2 P -M \cdot \left(s_{m_k}(x_i) + s_{m_k}(\bx_i)\right) &\le 2 P -
M/10  \\
&< -2P.
\end{align*}
The first expression uses $2P$ as the maximum possible payoff that player $j$
can obtain from the two other games in which he is involved. The first
inequality uses the fact that $i$ was the pair with maximal probability, and
there are exactly 10 pairs. The second inequality uses the fact that $M/10 >
4P$.

On the other hand, let $i'$ be an index in $J$. By the argument above, we have 
$s_{m_k}(x_{i'}) + s_{m_k}(\bx_{i'}) = 0$. Hence, the payoff of actions
$x_{i'}$ and $\bx_{i'}$ to player $j$ is at least $-2P$, since that is the
lowest payoff that he can obtain from the other two games in which he is
involved.

But now we have arrived at our contradiction. Player $j$ places non-zero
probability on at least one action $x_i$ or $\bx_i$ with $i \in I$ that is not
a pure best response. Hence $\sbf$ cannot be a Nash equilibrium.
\end{proof}

\myparagraph{\bf Gate gadgets.}
We now define the payoffs for variable and constraint players. Actions $x_i$
and $\bx_i$ of variable player $v_j$ will represent the output of gate
$g_{i,j}$. Specifically, the probability that player $v_j$ assigns to action
$x_i$ will be equal to the output of $g_{i, j}$. In this way, the strategy of
variable player $v_j$ will represent the output of every gate at
level $j$ of the circuit. The constraint player $c_j$ enforces all constraints
between the gates at level $j$ and the gates at level $j+1$. To simulate each
gate, we will embed one of the gate gadgets from \cref{fig:gadgets}, which
originated from the reduction of DGP~\cite{DGP}, into the bimatrix games that
involve $c_j$. 

%
\setboolean{main}{true}   
\begin{figure}
\begin{center}
\scalebox{0.49}{ 
\newsavebox\const
\begin{lrbox}{\const}
\ifthenelse{\boolean{main}}{
\begin{tikzpicture}
\matrix (m) [matrix of nodes, 
             	anchor=south west,
			 	ampersand replacement=\&,
				nodes={draw, minimum size=2cm},
				column sep={2cm,between origins},
                row sep={2cm,between origins},
             	nodes in empty cells,
             	inner sep=0pt
		]
{ 
\& \\
\& \\
};

\payoffs{1}{1}{$c$}{$0$}
\payoffs{1}{2}{$c$}{$1$}
\payoffs{2}{1}{$0$}{$1$}
\payoffs{2}{2}{$1$}{$0$}

\rowstrategy{1}{$\bar{x}_i$}
\rowstrategy{2}{$x_i$}

\colstrategy{1}{$\bar{x}_i$}
\colstrategy{2}{$x_i$}

\players{$c_j$}{$v_j$}

\end{tikzpicture}

}{
\begin{tikzpicture}
\matrix (m) [matrix of nodes, 
             	anchor=south west,
			 	ampersand replacement=\&,
				nodes={draw, minimum size=2cm},
				column sep={2cm,between origins},
                row sep={2cm,between origins},
             	nodes in empty cells,
             	inner sep=0pt
		]
{ 
\& \\
\& \\
};

\payoffs{1}{1}{$c$}{$0$}
\payoffs{1}{2}{$c$}{$1$}
\payoffs{2}{1}{$0$}{$1$}
\payoffs{2}{2}{$1$}{$0$}

\rowstrategy{1}{$\bar{x}_i$}
\rowstrategy{2}{$x_i$}

\colstrategy{1}{$\bar{x}_i$}
\colstrategy{2}{$x_i$}

\players{$c_j$}{$v_j$}

\end{tikzpicture}

}
\end{lrbox}

\newsavebox\add
\begin{lrbox}{\add}
\ifthenelse{\boolean{main}}{
\input \begin{tikzpicture}
\matrix (m) [matrix of nodes, 
             	anchor=south west,
			 	ampersand replacement=\&,
				nodes={draw, minimum size=2cm},
				column sep={2cm,between origins},
                row sep={2cm,between origins},
             	nodes in empty cells,
             	inner sep=0pt
		]
{ 
\& \& \& \& \& \\
\& \& \& \& \& \\
};

\payoffs{1}{1}{$0$}{$0$}
\payoffs{1}{2}{$1$}{$0$}
\payoffs{1}{3}{$0$}{$0$}
\payoffs{1}{4}{$1$}{$0$}
\payoffs{1}{5}{$0$}{$0$}
\payoffs{1}{6}{$0$}{$1$}
\payoffs{2}{1}{$0$}{$0$}
\payoffs{2}{2}{$0$}{$0$}
\payoffs{2}{3}{$0$}{$0$}
\payoffs{2}{4}{$0$}{$0$}
\payoffs{2}{5}{$0$}{$1$}
\payoffs{2}{6}{$1$}{$0$}

\colstrategy{1}{$\bar{x}_b$}
\colstrategy{2}{$x_b$}
\colstrategy{3}{$\bar{x}_c$}
\colstrategy{4}{$x_c$}
\colstrategy{5}{$\bar{x}_a$}
\colstrategy{6}{$x_a$}

\rowstrategy{1}{$\bar{x}_a$}
\rowstrategy{2}{$x_a$}

\players{$c_{j-1}$}{$v_{j-1}$}
\playersshift{}{$v_{j}$}{5}

\end{tikzpicture}
           
}{
\input \begin{tikzpicture}
\matrix (m) [matrix of nodes, 
             	anchor=south west,
			 	ampersand replacement=\&,
				nodes={draw, minimum size=2cm},
				column sep={2cm,between origins},
                row sep={2cm,between origins},
             	nodes in empty cells,
             	inner sep=0pt
		]
{ 
\& \& \& \& \& \\
\& \& \& \& \& \\
};

\payoffs{1}{1}{$0$}{$0$}
\payoffs{1}{2}{$1$}{$0$}
\payoffs{1}{3}{$0$}{$0$}
\payoffs{1}{4}{$1$}{$0$}
\payoffs{1}{5}{$0$}{$0$}
\payoffs{1}{6}{$0$}{$1$}
\payoffs{2}{1}{$0$}{$0$}
\payoffs{2}{2}{$0$}{$0$}
\payoffs{2}{3}{$0$}{$0$}
\payoffs{2}{4}{$0$}{$0$}
\payoffs{2}{5}{$0$}{$1$}
\payoffs{2}{6}{$1$}{$0$}

\colstrategy{1}{$\bar{x}_b$}
\colstrategy{2}{$x_b$}
\colstrategy{3}{$\bar{x}_c$}
\colstrategy{4}{$x_c$}
\colstrategy{5}{$\bar{x}_a$}
\colstrategy{6}{$x_a$}

\rowstrategy{1}{$\bar{x}_a$}
\rowstrategy{2}{$x_a$}

\players{$c_{j-1}$}{$v_{j-1}$}
\playersshift{}{$v_{j}$}{5}

\end{tikzpicture}
          
}
\end{lrbox}

\newsavebox\sub
\begin{lrbox}{\sub}
\ifthenelse{\boolean{main}}{
\input \begin{tikzpicture}
\matrix (m) [matrix of nodes, 
             	anchor=south west,
			 	ampersand replacement=\&,
				nodes={draw, minimum size=2cm},
				column sep={2cm,between origins},
                row sep={2cm,between origins},
             	nodes in empty cells,
             	inner sep=0pt
		]
{ 
\& \& \& \& \& \\
\& \& \& \& \& \\
};

\payoffs{1}{1}{$0$}{$0$}
\payoffs{1}{2}{$1$}{$0$}
\payoffs{1}{3}{$0$}{$0$}
\payoffs{1}{4}{$-1$}{$0$}
\payoffs{1}{5}{$0$}{$0$}
\payoffs{1}{6}{$0$}{$1$}
\payoffs{2}{1}{$0$}{$0$}
\payoffs{2}{2}{$0$}{$0$}
\payoffs{2}{3}{$0$}{$0$}
\payoffs{2}{4}{$0$}{$0$}
\payoffs{2}{5}{$0$}{$1$}
\payoffs{2}{6}{$1$}{$0$}

\colstrategy{1}{$\bar{x}_b$}
\colstrategy{2}{$x_b$}
\colstrategy{3}{$\bar{x}_c$}
\colstrategy{4}{$x_c$}
\colstrategy{5}{$\bar{x}_a$}
\colstrategy{6}{$x_a$}

\rowstrategy{1}{$\bar{x}_a$}
\rowstrategy{2}{$x_a$}

\players{$c_{j-1}$}{$v_{j-1}$}
\playersshift{}{$v_{j}$}{5}

\end{tikzpicture}
      
}{
\input \begin{tikzpicture}
\matrix (m) [matrix of nodes, 
             	anchor=south west,
			 	ampersand replacement=\&,
				nodes={draw, minimum size=2cm},
				column sep={2cm,between origins},
                row sep={2cm,between origins},
             	nodes in empty cells,
             	inner sep=0pt
		]
{ 
\& \& \& \& \& \\
\& \& \& \& \& \\
};

\payoffs{1}{1}{$0$}{$0$}
\payoffs{1}{2}{$1$}{$0$}
\payoffs{1}{3}{$0$}{$0$}
\payoffs{1}{4}{$-1$}{$0$}
\payoffs{1}{5}{$0$}{$0$}
\payoffs{1}{6}{$0$}{$1$}
\payoffs{2}{1}{$0$}{$0$}
\payoffs{2}{2}{$0$}{$0$}
\payoffs{2}{3}{$0$}{$0$}
\payoffs{2}{4}{$0$}{$0$}
\payoffs{2}{5}{$0$}{$1$}
\payoffs{2}{6}{$1$}{$0$}

\colstrategy{1}{$\bar{x}_b$}
\colstrategy{2}{$x_b$}
\colstrategy{3}{$\bar{x}_c$}
\colstrategy{4}{$x_c$}
\colstrategy{5}{$\bar{x}_a$}
\colstrategy{6}{$x_a$}

\rowstrategy{1}{$\bar{x}_a$}
\rowstrategy{2}{$x_a$}

\players{$c_{j-1}$}{$v_{j-1}$}
\playersshift{}{$v_{j}$}{5}

\end{tikzpicture}
      
}
\end{lrbox}

\newsavebox\mult
\begin{lrbox}{\mult}
\ifthenelse{\boolean{main}}{
\input \begin{tikzpicture}
\matrix (m) [matrix of nodes, 
             	anchor=south west,
			 	ampersand replacement=\&,
				nodes={draw, minimum size=2cm},
				column sep={2cm,between origins},
                row sep={2cm,between origins},
             	nodes in empty cells,
             	inner sep=0pt
		]
{ 
\& \& \& \\
\& \& \& \\
};

\payoffs{1}{1}{$0$}{$0$}
\payoffs{1}{2}{$c$}{$0$}
\payoffs{1}{3}{$0$}{$0$}
\payoffs{1}{4}{$0$}{$1$}
\payoffs{2}{1}{$0$}{$0$}
\payoffs{2}{2}{$0$}{$0$}
\payoffs{2}{3}{$0$}{$1$}
\payoffs{2}{4}{$1$}{$0$}

\colstrategy{1}{$\bar{x}_b$}
\colstrategy{2}{$x_b$}
\colstrategy{3}{$\bar{x}_c$}
\colstrategy{4}{$x_c$}

\rowstrategy{1}{$\bar{x}_a$}
\rowstrategy{2}{$x_a$}

\players{$c_{j/j-1}$}{$v_{j'}$}
\playersshift{}{$v_{j}$}{3}

\end{tikzpicture}

}{
\input \begin{tikzpicture}
\matrix (m) [matrix of nodes, 
             	anchor=south west,
			 	ampersand replacement=\&,
				nodes={draw, minimum size=2cm},
				column sep={2cm,between origins},
                row sep={2cm,between origins},
             	nodes in empty cells,
             	inner sep=0pt
		]
{ 
\& \& \& \\
\& \& \& \\
};

\payoffs{1}{1}{$0$}{$0$}
\payoffs{1}{2}{$c$}{$0$}
\payoffs{1}{3}{$0$}{$0$}
\payoffs{1}{4}{$0$}{$1$}
\payoffs{2}{1}{$0$}{$0$}
\payoffs{2}{2}{$0$}{$0$}
\payoffs{2}{3}{$0$}{$1$}
\payoffs{2}{4}{$1$}{$0$}

\colstrategy{1}{$\bar{x}_b$}
\colstrategy{2}{$x_b$}
\colstrategy{3}{$\bar{x}_c$}
\colstrategy{4}{$x_c$}

\rowstrategy{1}{$\bar{x}_a$}
\rowstrategy{2}{$x_a$}

\players{$c_{j/j-1}$}{$v_{j'}$}
\playersshift{}{$v_{j}$}{3}

\end{tikzpicture}

}
\end{lrbox}

\begin{tikzpicture}

\draw (-4, 7.5)  node {\usebox\add};
\node at (-3,4) {\LARGE $g_{a,j} = g_{b,j-1} +^b_{0.1} g_{c,j-1}$};

\draw (9, 7.5)  node {\usebox\mult};
\node at (10,4) {\LARGE $g_{a,j} = g_{b,j'} *^b_{0.1} c$};

\draw (-4, 0)   node {\usebox\sub};
\node at (-3, -3.5) {\LARGE $g_{a,j} = g_{b,j-1} -^b_{0.1} g_{c,j-1}$};

\draw (9, 0)  node {\usebox\const};
\node at (10,-3.5) {\LARGE $g_{i,j} = c$};

\end{tikzpicture} 

}
\caption{DGP polymatrix game gadgets.}
\label{fig:gadgets}
\end{center}
\end{figure}
%

The idea is that, for the constraint player to be in equilibrium, the
variable players must play $x_i$ with probabilities that exactly
simulate the original gate. 
\cref{lem:mixing} allows us to treat each gate independently: 
each pair of actions $x_i$ and $\sbf_i$ must receive probability $0.1$ in
total, but the split of probability between $x_i$ and $\sbf_i$ is
determined by the gate gadgets.

Formally, we construct the payoff matrices $A^{v_i, c_i}$ and $A^{c_i, v_{i+1}}$
for all $i < n$ by first setting each payoff to $0$. Then, for each gate, we
embed the corresponding gate gadget from \cref{fig:gadgets} into the
matrices. 
For each gate $g_{a,j}$, we take the corresponding game from
\cref{fig:gadgets}, and embed it into the rows $x_a$ and
$\bx_a$ of a constraint player's matrix. The diagrams specify
specific actions of the constraint and variable players that should be
modified. 

For gates that originated in the circuit, the gadget is always embedded into
the matrices $A^{v_{j-1}, c_{j-1}}$ and $A^{c_{j-1}, v_j}$, the synchronicity of
the circuit ensures that the inputs for level $j$ gates come from level $j-1$
gates. We have also added extra multiplication gates that copy values from the
output of the circuit back to the input.  These gates are of the form $g_{i,j} =
g_{i', j+1}$, and are embedded into the matrices $A^{v_{j}, c_{j}}$ and
$A^{c_{j}, v_{j+1}}$.

The following lemma, states that, in
every Nash equilibrium, the
strategies of the variable players exactly simulate the gates that have been
embedded.

\begin{lemma}
\label{lem:simulation}
In every mixed Nash equilibrium $\sbf$ of the game, 
the following are satisfied for each gate $g_{i,j}$.
\begin{itemize}
\item If $g_{i,j} = c$, then $s_{v_j}(x_i) = c$.

\item If $g_{i,j} = g_{i_1, j-1} \; \badd_{0.1} \; g_{i_2, j-1}$, then $s_{v_j}(x_i) = s_{v_{j-1}}(x_{i_1}) \; \badd_{0.1} \; s_{v_{j-1}}(x_{i_2})$.

\item If $g_{i,j} = g_{i_1, j-1} \; \bsub_{0.1} \; g_{i_2, j-1}$, then
$s_{v_{j}}(x_i) =
s_{v_{j-1}}(x_{i_1}) \; \bsub_{0.1} \; s_{v_{j-1}}(x_{i_2})$.

\item If $g_{i,j} = g_{i_1, j'} \; \bmul_{0.1} \; c$, then $s_{v_j}(x_i) =
s_{v_{j'}}(x_{i_1}) \; \bmul_{0.1} \; c$.
\end{itemize}
\end{lemma}
\begin{proof}
We can actually prove this lemma for all four gates simultaneously. Let $j'$ be
the index constraint player into which the gate gadget is embedded. Observe that
all four games for the four gate types have a similar structure: The payoffs for actions $x_i$
and $\bx_i$ for player $v_j$ are identical across all four games, and the payoff
of action $x_i$ for $c_{j'}$ are also identical;
the only thing that differs between the gates is the payoff to player $c_{j'}$
for action~$\bx_i$. We describe these differences using a function $f$.
\begin{itemize}
\item For $c$-gates, we define $f(\sbf) = c$.
\item For $\badd_{0.1}$-gates, we define $f(\sbf) = s_{v_{j-1}}(x_{i_1}) \; + \; s_{v_{j-1}}(x_{i_1})$.

\item For $\bsub_{0.1}$-gates, we define $f(\sbf) = s_{v_{j-1}}(x_{i_1}) \; - \; s_{v_{j-1}}(x_{i_1})$.

\item For $\bmul_{0.1}$-gates, we define $f(\sbf) = s_{v_{j'}}(x_{i_1}) \;
* \; c$.
\end{itemize}
Observe that the payoff of action $\bx_i$ to player $c_{j'}$ is $f(\sbf)$. 
To prove the lemma, we must show that player $v_j$ plays $x_i$ with probability
$$\min(\max(f(\sbf), 0.1), 0).$$ There are three cases to consider.
\begin{itemize}
\item If $f(\sbf) \le 0$, then we argue that $s_{v_j}(x_i) = 0$.
Suppose for the sake of contradiction that player $v_j$ places non-zero
probability on action $x_i$. Then action $x_i$ for player $c_{j'}$ will have
payoff strictly greater than zero,
whereas action $\bx_i$ will have payoff $f(\sbf) \le 0$. Hence, in
equilibrium, $c_{j'}$ cannot play action $\bx_i$. \cref{lem:mixing} then
implies that player $c_{j'}$ must play $x_i$ with probability $0.1$.  If $c_{j'}$ does
this, then the payoff to $v_j$ for $x_i$ will be zero, and the payoff to $v_j$
for $\bx_i$ will be $0.1$. This means that $v_j$ places non-zero probability on
an action that is not a best response, and so is a contradiction.

\item If $f(\sbf) \ge 0.1$, then we argue that $s_{v_j}(x_i) = 0.1$. Suppose
for the sake of contradiction with \cref{lem:mixing} that $s_{v_j}(\bx_i) > 0$. 
Observe that the payoff
to player $c_{j'}$ of action $\bx_i$ is $f(\sbf) \ge 0.1$, whereas the payoff to
player $c_{j'}$ of action $x_i$ is $s_{v_j}(x_i) < 0.1$. So to be in
equilibrium and consistent with \cref{lem:mixing},
player $c_{j'}$ must place $0.1$ probability on action $\bx_i$, and $0$
probability on action $x_i$. But this means that the payoff of action~$\bx_i$ 
to player $v_j$ is zero, while the payoff of action $x_i$ to player
$v_j$ is $0.1$. Hence player $v_j$ has placed non-zero probability on an action
that is not a pure best response, and so we have our contradiction.

\item If $0 < f(\sbf) < 0.1$, then we argue that $s_{v_j}(x_i) = f(\sbf)$. We
first prove that player $c_{j'}$ must play both $x_i$ and $\bx_i$ with positive
probability.
\begin{itemize}
\item If player $c_{j'}$ does not play $\bx_i$ then player $v_j$ will not
play $x_i$, and player $c_{j'}$ will receive payoff $0$, but in this scenario he
could get $f(\sbf) > 0$ by playing $\bx_i$ instead of his current strategy.
\item 
If player $c_{j'}$ does not play $x_i$ then player $v_j$ will not play $\bx_i$.
Player $c_{j'}$ will receive payoff $f(\sbf)$ for playing $\bx_i$, but in this
scenario he could receive payoff $1 > f(\sbf)$ for playing $x_i$ instead.
\end{itemize}
In order for player $c_{j'}$ to mix over $x_i$ and $\bx_i$ in equilibrium, their
payoffs must be equal. This is only the case when $s_{v_j}(x_i) = f(\sbf)$.
\end{itemize}
\end{proof}

\cref{lem:simulation} says that, in every Nash equilibrium of the game, the
strategies of the variable players exactly simulate the gates, which by
construction means that they give us a fixed point of the circuit $F$.
Also note that it is straightforward to give a path decomposition for our
interaction graph, where each node in the decomposition contains exactly two
vertices from the game, meaning that the graph has pathwidth~1.
So we have proved the following.

\begin{theorem}
\label{thm:caterpillar}
It is \PPAD-hard to find a Nash equilibrium of a tree polymatrix game, even when
all players have at most twenty actions and the interaction graph has
pathwidth~1.
\end{theorem}


\section{\PPAD hardness for paths}
In this section we show that we can extend our \PPAD-hardness result to
polymatrix games played on paths. We prove this by giving a generic reduction
from a polymatrix game played on a caterpillar graph to a polymatrix game
played on a path such that the Nash equilibria of the path game correspond to
the Nash equilibria of the original caterpillar game. 

\begin{figure}
\begin{center}
\scalebox{0.7}{ 
\begin{tikzpicture}
\node [state]             (v1) {$v_1$};
\node [state,below of=v1] (m1) {$m_1$};

\node [state,right of=v1] (c1) {$v_2$};
\node [state,below of=c1] (m2) {$m_2$};

\node [state,right of=c1] (v2) {$v_3$};
\node [state,below of=v2] (m3) {$m_3$};

\node [state,right of=v2] (c2) {$v_4$};
\node [state,below of=c2] (m4) {$m_4$};

\node [state,right of=c2] (v3) {$v_5$};
\node [state,below of=v3] (m5) {$m_5$};

\node [right of=v3,node distance=2cm,minimum size=2cm] (mid) {$\dots$};

\node [state,right of=v3,node distance=4cm] (vn) {$v_n$};
\node [state,below of=vn] (mn) {$m_{n}$};

\path 
    (v1) edge (m1)
    (c1) edge (m2)
    (v2) edge (m3)
    (c2) edge (m4)
    (v3) edge (m5)
    (vn) edge (mn)

    (v1) edge (c1)
    (c1) edge (v2)
    (v2) edge (c2)
    (c2) edge (v3)

    (v3) edge (mid)
    (mid) edge (vn)
    ;

\end{tikzpicture}
}
\end{center}
\caption{The input of our path polymatrix reduction.}
\label{fig:pathinput}
\end{figure}

For the purposes of this section, we assume that we are given a polymatrix
game~$\mathcal{T}$ of the form shown in Figure~\ref{fig:pathinput}. That is,
our caterpillar graph consists of a path of players $v_1$ through $v_n$ where each
player $v_i$ has an edge to a player $m_{i}$. In particular, when compared to
the previous section, we now drop the distinction between the $v_i$ players and the
$c_i$ players, since the reduction that we present here can be applied to any
polymatrix game that is played on a caterpillar graph.

Before we apply the reduction, it will be technically convenient to first
modify the game by adding three extra dummy players.
\begin{itemize}
\item We add a player $v_0$ with an edge to $v_1$. 
\item We add a player $v_{n+1}$ with an edge to $v_{n}$. 
\item We add a player $m_{n+1}$ with an edge to $v_{n+1}$. 
\end{itemize}
All of the payoff matrices of the newly introduced players are zero matrices,
so they do not strategically interact with the game. Let $\mathcal{T'}$ be the
game with the three new players added.

\begin{figure}
\begin{center}
\scalebox{0.7}{ 
\begin{tikzpicture}
\node [state]             (v1) {$v_1$};
\node [state,below of=v1] (m1) {$m_1$};

\node [state,right of=v1] (c1) {$v_2$};
\node [state,below of=c1] (m2) {$m_2$};

\node [state,right of=c1] (v2) {$v_3$};
\node [state,below of=v2] (m3) {$m_3$};

\node [state,right of=v2] (c2) {$v_4$};
\node [state,below of=c2] (m4) {$m_4$};

\node [state,right of=c2] (v3) {$v_5$};
\node [state,below of=v3] (m5) {$m_5$};

\node [state,left of=v1] (v0) {$v_0$};

\node [right of=v3,node distance=2cm,minimum size=2cm] (mid) {$\dots$};

\node [state,right of=v3,node distance=4cm] (vn) {$v_n$};
\node [state,below of=vn] (mn) {$m_{n}$};

\node [state,right of=vn] (vnp) {$v_{n+1}$};
\node [state,below of=vnp] (mnp) {$m_{n+1}$};

\path 
    (v0) edge (v1)
    (v1) edge (m1)
    (c1) edge (m2)
    (v2) edge (m3)
    (c2) edge (m4)
    (v3) edge (m5)
    (vn) edge (mn)
    (vn) edge (vnp)
    (vnp) edge (mnp)

    (v1) edge (c1)
    (c1) edge (v2)
    (v2) edge (c2)
    (c2) edge (v3)

    (v3) edge (mid)
    (mid) edge (vn)

    (v0) edge [ultra thick, red, <->] node [below left] {$T_0$} (m1)
    (v1) edge [ultra thick, red, <->] node [below left] {$T_1$} (m2)
    (c1) edge [ultra thick, red, <->] node [below left] {$T_2$} (m3)
    (v2) edge [ultra thick, red, <->] node [below left] {$T_3$} (m4)
    (c2) edge [ultra thick, red, <->] node [below left] {$T_4$} (m5)
    (vn) edge [ultra thick, red, <->] node [below left] {$T_n$} (mnp)
    ;

\end{tikzpicture}
}
\end{center}
\caption{The structure of the simulation used by $\mathcal{P(T')}$.}
\label{fig:pathreduction}
\end{figure}

Then we reduce the game $\mathcal{T'}$ to the path polymatrix game
$\mathcal{P(T')}$. The game $\mathcal{P(T')}$ contains the players $T_0$,
$T_1$, $\dots$, $T_n$ who are arranged in a path, meaning that each player
$T_i$ with $0 < i < n$ has an edge to $T_{i-1}$ and to $T_{i+1}$. The idea is
that player~$T_i$ is responsible for simulating players $v_i$ and $m_{i+1}$
from the game $T'$. This is shown in Figure~\ref{fig:pathreduction}, where the
red arrows show the players that each $T_i$ player is simulating.

Each player $T_i$ will have a strategy space that consists of all possible
pairs of actions for the two players that $T_i$ is simulating. 
Formally, for each pair of
actions $\lambda^i_{j}$ and $\mu^{i+1}_{k}$ of $v_i$ and $m_{i+1}$,
respectively, we create the action
$(\lambda^i_{j}, \mu^{i+1}_{k})$ for player $T_i$. Hence, player $T_i$ has
at most $\ell^2$ actions, where $\ell$ is the maximum number of actions used by
any player in $\mathcal{T}$. 

When player $T_i$ chooses the action $(\lambda^i_{j}, \mu^{i+1}_{k})$, the
payoffs are defined as follows.
\begin{itemize}
    \item When player $T_{i-1}$ chooses action 
    $(\lambda^{i-1}_{j'}, \mu^i_{k'})$, 
    the payoff to player $T_i$ from this interaction is 
    $A^{v_i,v_{i-1}}(\lambda^i_j,\lambda^{i-1}_{j'})+ A^{v_i,m_i}(\lambda^i_j, \mu^i_{k'})$.
    \item When player $T_{i+1}$ chooses action 
    $(\lambda^{i+1}_{\hat{j}}, \mu^{i+2}_{\hat{k}})$, 
    the payoff to player $T_i$ from this interaction is 
    $A^{v_i,v_{i+1}}(\lambda^i_j,\lambda^{i+1}_{\hat{j}})+ A^{m_{i+1},v_{i+1}}(\mu^{i+1}_k,\lambda^{i+1}_{\hat{j}})$.
\end{itemize}

The following lemma shows that  $\mathcal{P(T')}$ correctly simulates
$\mathcal{T}$. 

\begin{lemma}
    \label{lem:caterpillar_to_path}
    Given a Nash equilibrium of $\mathcal{P(T')}$, we can construct in polynomial time a Nash equilibrium for $\mathcal{T}$.
\end{lemma}
\begin{proof}
    Assume that $s$ is a Nash equilibrium for $\mathcal{P(T')}$, where player
$T_i$ plays strategy $s_{T_i}$. We construct a strategy profile for
$\mathcal{T}$ as follows.
    \begin{itemize}
        \item Player $v_i$ plays action $\lambda^i_j$ with probability 
        $s'_{v_i}(\lambda^i_j) := \sum_{k} s_{T_i}(\lambda^i_j, \mu^{i+1}_k)$.
        \item Player $m_i$ plays action $\mu^i_k$ with probability 
        $s'_{m_i}(\mu^{i}_k) := \sum_{j} s_{T_{i-1}}(\lambda^{i-1}_j,\mu^i_k)$.
    \end{itemize}
    Observe that by the construction of $s'$ we get that 
    \begin{itemize}
        \item $s'_{v_i}(\lambda^i_j) > 0$ if and only if $s_{T_i} (\lambda^i_j, \mu^{i+1}_k) > 0$ 
    and that 
    \item $s'_{m_{i+1}}(\mu^{i+1}_k) > 0$ if and only if $s_{T_i} (\lambda^i_j, \mu^{i+1}_k) > 0$.
    \end{itemize}

    In what follows, for notation simplicity, for any two players $x,y$ of the
tree polymatrix game $\mathcal{T}$, any action $j$ of player $x$, and any
strategy profile $s'$ let us denote $u_{(x,y)}(j,s')$ the expected payoff
player $x$ gets from action $j$ from player $y$ when player $y$ plays according
to $s'$.
    
    Now, let us compute the payoff player $T_i$ gets from action $(\lambda^i_j,
\mu^{i+1}_k)$ under strategy profile $s$. From the interaction with player
$T_{i-1}$ the player gets
    \begin{align*}
        & \sum_{(j',k')} s_{T_{i-1}} (\lambda^{i-1}_{j'}, \mu^i_{k'}) \cdot \left [ A^{v_i,v_{i-1}}(\lambda^i_j,\lambda^{i-1}_{j'})+ A^{v_i,m_i}(\lambda^i_j, \mu^i_{k'}) \right ] \\
       = & \sum_{(j', k')} s_{T_{i-1}} (\lambda^{i-1}_{j'}, \mu^i_{k'}) \cdot A^{v_i,v_{i-1}}(\lambda^i_j,\lambda^{i-1}_{j'}) + \sum_{(j',k')} s_{T_{i-1}} (\lambda^{i-1}_{j'}, \mu^i_{k'}) \cdot A^{v_i,m_i}(\lambda^i_j, \mu^i_{k'}) \\
        = & \sum_{j'} \sum_{k'} s_{T_{i-1}} (\lambda^{i-1}_{j'}, \mu^i_{k'}) \cdot A^{v_i,v_{i-1}}(\lambda^i_j,\lambda^{i-1}_{j'}) +
       \sum_{k'} \sum_{j'} s_{T_{i-1}} (\lambda^{i-1}_{j'}, \mu^i_{k'}) \cdot A^{v_i,m_i}(\lambda^i_j, \mu^i_{k'}) \\
       = & \sum_{j'} s_{v_{i-1}} (\lambda^{i-1}_{j'}) \cdot A^{v_i,v_{i-1}}(\lambda^i_j,\lambda^{i-1}_{j'}) +
       \sum_{k'} s_{m_{i}} (\mu^i_{k'}) \cdot A^{v_i,m_i}(\lambda^i_j, \mu^i_{k'})\\
       = & u_{(v_i,v_{i-1})}({\lambda^i_j},s')+ u_{(v_i,m_{i})}({\lambda^i_j}, s').
    \end{align*}
    From the interaction with player $T_{i+1}$ the player gets
    \begin{align*}
        & \sum_{(\hat{j},\hat{k})} s_{T_{i+1}} (\lambda^{i+1}_{\hat{j}}, \mu^{i+2}_{\hat{k}}) \cdot \left [ A^{v_i,v_{i+1}}(\lambda^i_j,\lambda^{i+1}_{\hat{j}}) + 
        A^{m_{i+1},v_{i+1}}(\mu^{i+1}_k,\lambda^{i+1}_{\hat{j}}) \right ] \\
       = & \sum_{(\hat{j},\hat{k})} s_{T_{i+1}} (\lambda^{i+1}_{\hat{j}}, \mu^{i+2}_{\hat{k}}) \cdot A^{v_i,v_{i+1}}(\lambda^i_j,\lambda^{i+1}_{\hat{j}}) + 
        \sum_{(\hat{j},\hat{k})} s_{T_{i+1}} (\lambda^{i+1}_{\hat{j}}, \mu^{i+2}_{\hat{k}}) \cdot A^{m_{i+1},v_{i+1}}(\mu^{i+1}_k,\lambda^{i+1}_{\hat{j}})  \\
        = & \sum_{\hat{j}}\sum_{\hat{k}} s_{T_{i+1}} (\lambda^{i+1}_{\hat{j}}, \mu^{i+2}_{\hat{k}}) \cdot A^{v_i,v_{i+1}}(\lambda^i_j,\lambda^{i+1}_{\hat{j}}) + 
        \sum_{\hat{k}}\sum_{\hat{j}}s_{T_{i+1}} (\lambda^{i+1}_{\hat{j}}, \mu^{i+2}_{\hat{k}}) \cdot A^{m_{i+1},v_{i+1}}(\mu^{i+1}_k,\lambda^{i+1}_{\hat{j}})  \\
        = & \sum_{\hat{j}} s_{v_{i+1}} (\lambda^{i+1}_{\hat{j}}) \cdot A^{v_i,v_{i+1}}(\lambda^i_j,\lambda^{i+1}_{\hat{j}}) + 
        \sum_{\hat{k}} s_{m_{i+1}}(\mu^{i+2}_{\hat{k}}) \cdot A^{m_{i+1},v_{i+1}}(\mu^{i+1}_k,\lambda^{i+1}_{\hat{j}})\\
        = & u_{(v_i,v_{i+1})}({\lambda^i_j}, s')+ u_{(m_{i+1},v_{i+1})}({\mu^{i+1}_k},s').
    \end{align*}

    Since strategy profile $s$ is a Nash equilibrium of $\mathcal{P(T')}$, we
know that player $T_i$ puts positive probability on actions $(\lambda^i_j,
\mu^{i+1}_k)$ with maximum payoff. Hence, if we combine the calculations above,
we conclude that when $s_{T_i}(\lambda^i_j, \mu^{i+1}_k) > 0$
    \begin{align}
        \nonumber
        u_{(v_i,v_{i-1})}({\lambda^i_j},s')+ u_{(v_i,m_{i})}({\lambda^i_j}, s') + u_{(v_i,v_{i+1})}({\lambda^i_j}, s')+ u_{(m_{i+1},v_{i+1})}({\mu^{i+1}_k},s')  
        = \\
        \nonumber
        \max_{\lambda^i_{j'}, \mu^{i+1}_{k'}} \Big (
        u_{(v_i,v_{i-1})}({\lambda^i_{j'}},s')+ u_{(v_i,m_{i})}({\lambda^i_{j'}}, s') + u_{(v_i,v_{i+1})}({\lambda^i_{j'}},s')+ u_{(m_{i+1},v_{i+1})}(\mu^{i+1}_{k'},s') 
        \Big) = \\
        \label{eq:path_in_support}
        \max_{\lambda^i_{j'}} \Big (
        u_{(v_i,v_{i-1})}({\lambda^i_{j'}},s')+ u_{(v_i,m_{i})}({\lambda^i_{j'}}, s') + u_{(v_i,v_{i+1})}({\lambda^i_{j'}}, s')\Big) +
        \max_{\mu^{i+1}_{k'}} \Big (u_{(m_{i+1},v_{i+1})}({\mu^{i+1}_{k'}},s') \Big)   .
    \end{align}
    
We are ready to prove that the constructed strategy profile $s'$ is a Nash
equilibrium for $\mathcal{T}$. For the sake of contradiction, assume that $s'$
is not a Nash equilibrium. We will consider two cases, depending on which
player has incentives to deviate.
    \begin{itemize}
        \item Player $v_i$ wants to deviate. This means that there is an action $\lambda^i_{j_1}$ such that $s'_{v_i}(\lambda^i_{j_1}) > 0$, but it does not achieve maximum payoff for player $v_i$. This means that that there exists a $j'$ such that 
        \begin{align*}
            u_{(v_i,v_{i-1})}({\lambda^i_j},s')+ u_{(v_i,m_{i})}({\lambda^i_j}, s') + u_{(v_i,v_{i+1})}({\lambda^i_j}, s') < \\
            u_{(v_i,v_{i-1})}({\lambda^i_{j'}},s')+ u_{(v_i,m_{i})}({\lambda^i_{j'}}, s') + u_{(v_i,v_{i+1})}({\lambda^i_{j'}}, s').
        \end{align*}
        But this in turn means that player $T_i$ places positive probability of action $(\lambda^i_j, \mu^{i+1}_k)$ but there is an action $(\lambda^i_{j'}, \mu^{i+1}_k)$ that  Equation~\eqref{eq:path_in_support} is not satisfied. Hence we get a contradiction. 
        \item Player $m_{i+1}$ wants to deviate. This means that there is an action $\mu^{i+1}_k$ such that $s'_{m_{i+1}}(\mu^{i+1}_k) > 0$, but it does not achieve maximum payoff for player $m_{i+1}$. Hence there is a $k'$ such that 
        \begin{align*}
            u_{(m_{i+1},v_{i+1})}({\mu^{i+1}_k},s') < u_{(m_{i+1},v_{i+1})}({\mu^{i+1}_{k'}},s').
        \end{align*}
        Again this means that player $T_i$ places positive probability of action $(\lambda^i_j, \mu^{i+1}_k)$ but there is an action $(\lambda^i_{j}, \mu^{i+1}_{k'})$ that  Equation~\eqref{eq:path_in_support} is not satisfied. Hence we get a contradiction.
    \end{itemize}    
\end{proof}

So we can apply the reduction to the family of hard instances given by
Theorem~\ref{thm:caterpillar} to obtain the following result, where we recall
that the instances used in 
Theorem~\ref{thm:caterpillar} have at most twenty actions per player, and that
this reduction at most squares the number of actions from the input game. 

\begin{theorem}
    \label{thm:paths-hard}
    It is \PPAD-complete to find a Nash equilibrium of a path polymatrix game, even when every player has at most four hundred actions.
\end{theorem}

\section{Conclusions and Future Work}
In this paper we have identified the intractability boundary of two problems.
\begin{itemize}
    \item For tree polymatrix games, we have shown that it is \PPAD-complete to
find an exact Nash equilibrium even when every player has at most twenty actions. This
complements the approximation schemes that exist for the problem~\cite{BLP,OI}. 
    \item For path polymatrix games, we have shown that it is \PPAD-complete to
find an exact Nash equilibrium even when every player has at most four hundred actions.
    \item It is \PPAD-complete to find an $\eps$-fixed point of a \twoDlinFIXP instance for all $\eps < 0.5$; while for $\eps=0.5$ the problem is trivial.
\end{itemize}

We conclude with some open questions that deserve further investigation. On a
practical level, the FPTAS of~\cite{OI} for $\eps$-NE in tree polymatrix games
reduces the problem to a dynamic programming approach that in each step it
solves a rather complicated CSP. Is there a more practical and intuitive
algorithm for the problem? From the theoretical point of view, the exact
boundary between tractability and hardness in tree polymatrix games is an
intriguing open question. As we show in this paper, twenty-action pathwidth-1
tree polymatrix games are hard, and four hundred action path polymatrix games
are hard, but two-action path polymatrix games can be solved in polynomial
time~\cite{EGG}. What about two-action tree polymatrix games, or
path-polymatrix games with more than two actions? Both problems seem quite
challenging and we expect that new techniques will be required in order to
tackle them.

\section*{Acknowledgments}
We would like to thank Alexandros Hollender for useful discussions that led to
Theorem~\ref{thm:linfixp-tight}. We would also like to thank an anonymous
reviewer for pointing out that our hardness result for caterpillar polymatrix
games implies the same hardness result for path polymatrix games.

\bibliographystyle{siamplain}
\bibliography{references}
\end{document}

%% file: ex_shared.tex
\usepackage{lipsum}
\usepackage{amsfonts}
\usepackage{graphicx}
\usepackage{epstopdf}
\usepackage{algorithmic}
\ifpdf
  \DeclareGraphicsExtensions{.eps,.pdf,.png,.jpg}
\else
  \DeclareGraphicsExtensions{.eps}
\fi

\newsiamremark{remark}{Remark}
\newsiamremark{hypothesis}{Hypothesis}
\crefname{hypothesis}{Hypothesis}{Hypotheses}
\newsiamthm{claim}{Claim}

\headers{Path Polymatrix Games are PPAD-hard}{A. Deligkas, J. Fearnley, and R. Savani}

\title{Path Polymatrix Games are PPAD-hard\thanks{A conference version of this paper has appeared at ICALP 2020. This version contains full proofs, a new improved and tight lower bound in Theorem 3.7, and also extends the hardness result to polymatrix games played on paths.
\funding{This work was funded by by EPSRC Grant EP/X039862/1 ``NAfANE: New Approaches for Approximate Nash Equilibria'' and EPSRC Grant EP/W014750/1
``New Techniques for Resolving Boundary Problems in Total Search''.}}}

\author{Argyrios Deligkas\thanks{Royal Holloway University of London, UK
  (\email{argyrios.deligkas@rhul.ac.uk}).}
\and John Fearnley\thanks{University of Liverpool, UK 
  (\email{john.fearnley@liverpool.ac.uk}, \email{rahul.savani@liverpool.ac.uk}).}
\and Rahul Savani\footnotemark[3]}

\usepackage{amsopn}


%% file: macros.tex
\usepackage{ifthen}
\newboolean{main}   

\usepackage{tikz}
\usepackage{pgfplots}
\usepackage{pgfplotstable}
\usetikzlibrary{matrix,arrows,shapes,positioning,calc,snakes,decorations.markings}
\usetikzlibrary{patterns}

\usetikzlibrary{arrows.meta}
\tikzset{>={Latex[width=2.5mm,length=2mm]}}

\pgfdeclarelayer{background}
\pgfdeclarelayer{foreground}
\pgfsetlayers{background,main,foreground}

\definecolor{lavander}{cmyk}{0,0.48,0,0}
\definecolor{violet}{cmyk}{0.79,0.88,0,0}
\definecolor{burntorange}{cmyk}{0,0.52,1,0}

\def\oran{orange!30}

\tikzstyle{players}=[draw,circle,burntorange, left color=\oran,
                       text=violet,minimum width=1cm]

\tikzstyle{playerssmall}=[draw,circle, minimum width=0.5cm]

\tikzstyle{gate}=[draw,circle,minimum width=1.25cm]

\tikzstyle{rect}=[rectangle] 

\pgfmathsetmacro{\os}{0.5}

\newcommand{\payoffs}[4]{ 
	\node[xshift=\os cm,yshift=\os cm] at (m-#1-#2.south west) {\LARGE #3};
	\node[xshift=-\os cm,yshift=-\os cm] at (m-#1-#2.north east) {\LARGE #4};
}
\newcommand{\rowstrategy}[2]{ 
	\node[xshift=-\os cm] at (m-#1-1.west) {\LARGE #2};
}
\newcommand{\colstrategy}[2]{ 
	\node[yshift=\os cm] at (m-1-#1.north) {\LARGE #2};
}
\newcommand{\playersshift}[3]{ 
	\coordinate (tl1) at (m-1-#3.north west);
	\coordinate (tl2) at ($(tl1)+(-1,1.25)$);
	\coordinate (tl3) at ($(tl1)!0.5!(tl2)$);
	\draw[-] (tl1) -- (tl2);
	\node[xshift=-0.2cm, yshift=0.1cm, anchor=east] at (tl3) {\LARGE #1};
	\node[xshift=0.2cm, yshift=0.45cm] at (tl3) {\LARGE #2};
}
\newcommand{\players}[2]{ 
	\playersshift{#1}{#2}{1}
}